\documentclass[11pt, twoside]{article}
\usepackage{atmp}
\copyrightnotice{2002}{6}{357}{401}
\usepackage{graphicx}
\newtheorem{theorem}{Theorem}

\newtheorem{lemma}[theorem]{Lemma}

\newtheorem{proposition}[theorem]{Proposition}
\newtheorem{remark}[theorem]{Remark}

\input{tcilatex}

\begin{document}

\title{
{\Large \textbf{\ Conformal modes in simplicial quantum gravity 
and the Weil-Petersson volume of moduli space }}}
\url{math-ph/0107028}

\author{M. Carfora, A. Marzuoli}
\address{Universit\`{a} degli Studi di Pavia, \\
via A. Bassi 6, I-27100 Pavia, Italy, \\
and\\
Istituto Nazionale di Fisica Nucleare, Sezione di Pavia, \\
via A. Bassi 6, I-27100 Pavia, Italy}
\addressemail{mauro.carfora@pv.infn.it}
\addressemail{annalisa.marzuoli@pv.infn.it}

\markboth{\it Conformal modes in simplicial quantum gravity\ldots}{\it M. Carfora, A. Marzuoli}
\begin{abstract}
Our goal here is to present a detailed analysis connecting the anomalous
scaling properties of 2D simplicial quantum gravity to the geometry of \ the
moduli space $\overline{\mathfrak{M}}_{g},_{N_{0}}$ of genus $g$ Riemann
surfaces with $N_{0}$ punctures.\ In the case of pure gravity we prove that
the scaling properties of the set of dynamical triangulations with $N_{0}$
vertices are directly provided by the large $N_{0}$ asymptotics of the
Weil-Petersson volume of $\overline{\mathfrak{M}}_{g},_{N_{0}}$, recently
discussed by Manin and Zograf. Such a geometrical characterization explains
why dynamical triangulations automatically take into account the anomalous
scaling properties of Liouville theory. In the case of coupling with
conformal matter we briefly argue that the anomalous scaling of the
resulting discretized\ theory should be related to the\ Gromov-Witten
invariants of the moduli space $\overline{\mathfrak{M}}_{g},_{N_{0}}(X,\beta )$
of stable maps from (punctured Riemann surfaces associated with) dynamical
triangulations to a (smooth projective) manifold $X$ parameterizing the
conformal matter configurations.\ 
\end{abstract}

\newpage

\section{\protect\bigskip Introduction}

The starting point of the path quantization of 2D gravity is the observation
that the space of Riemannian structures $Riem(M)$ on a $2$-dimensional
Riemannian manifold $M$ of genus $g$ can be decomposed by means of a local
slice for the action of \ the group of confeomorphisms\ $W(M)\ltimes Diff(M)$%
. This yields for the possibility of parameterizing locally a generic metric 
$h\in Riem(M)$ by means of a diffeomorphism $f\in Diff(M)$, a Weyl rescaling 
$e^{2u}\in W(M)$, and $3g-3$ complex parameters $\{m_{\alpha }\}$ varying in
the moduli space $\mathfrak{M}_{g}$ of genus $g$ Riemann surfaces. Explicitly we
can write 
\begin{gather}
S:\mathfrak{M}_{g}\times \left( W(M)\ltimes Diff(M)\right) \rightarrow Riem(M)
\label{meslice} \\
(\widehat{h}_{ab}(\mu _{\alpha }),e^{2v},\psi )\longmapsto S(\widehat{h}%
_{ab}(\mu _{\alpha }),e^{2v},\psi )\doteq h_{ab}=e^{2v}\left( \psi ^{\ast }%
\widehat{h}(m_{\alpha })\right) _{ab},  \notag
\end{gather}
where $\widehat{h}_{ab}(m_{\alpha })$ are the components of the reference
metric whose conformal class defines the point $\{m_{\alpha }\}$\ in $\mathfrak{M%
}_{g}$ we are considering. A natural choice \ for \ $\widehat{h}%
_{ab}(m_{\alpha })$ is associated with the slice defined by the metrics of
constant curvature. Such a framework allows us to introduce formal
functional measures $D\left[ h\right] $, $D[\psi ]$, and $D_{h}[v]$
respectively over (the tangent spaces to) \ $Riem(M)$, $Diff(M)$ and $W(M)$,
and whose properties, under conformal transformations, are instrumental to
the theory. To briefly review the main features of such an analysis, (see 
\emph{e.g.} [1]), let us introduce on \ $M$\ \ local complex coordinates $(z,%
\overline{z})$ in which $h=2h_{z\overline{z}}|dz|^{2}$ and define 
\begin{equation}
\begin{tabular}{ccc}
$\nabla _{(n)}^{z}\phi $ & $=$ & $(h_{z\overline{z}})^{-1}\frac{\partial }{%
\partial \overline{z}}\phi \otimes (dz)^{-1}$ \\ 
$\nabla _{z}^{(n)}\phi $ & $=$ & $(h_{z\overline{z}})^{n}\frac{\partial }{%
\partial z}(h_{z\overline{z}})^{-n}\phi \otimes dz$ \\ 
$\Delta _{(n)}^{+}$ & $=$ & $-2\nabla _{(n+1)}^{z}\nabla _{z}^{(n)}$ \\ 
$\Delta _{(n)}^{-}$ & $=$ & $-2\nabla _{z}^{(n-1)}\nabla _{(n)}^{z}$%
\end{tabular}
,
\end{equation}
where $\phi =\phi _{z...z}(dz)^{n}$\ is a tensor field of weight $(n,0)$.
The conformal properties of \ $D\left[ h\right] $\ are strictly connected to
the behavior of the family of functional determinants (arising as jacobians
of the slice map (\ref{meslice})) 
\begin{equation}
Z_{(n)}^{\pm }(h)\equiv \frac{\det \,^{\prime }\Delta _{(n)}^{\pm }}{\det
\left\langle \phi _{j}\right| \left. \phi _{r}\right\rangle _{h}\det
\left\langle \psi _{a}\right| \left. \psi _{b}\right\rangle _{h}}.
\end{equation}
where $\phi _{j}\in \ker \nabla _{(n+1)}^{z}$, $\psi _{a}\in \ker \nabla
_{z}^{(n)}$ for $Z_{(n)}^{+}(h)$, $\phi _{j}\in \ker \nabla
_{z}^{(n-1)}$, $\psi _{a}\in \ker \nabla _{(n)}^{z}$ for $Z_{(n)}^{-}(h)$,
and where $\det \,^{\prime }\Delta _{(n)}^{\pm }$ denotes the $\zeta $%
-regularized determinant restricted to the non-zero modes. According to a
well-known result, the behavior of \ $Z_{(n)}^{\pm }(h)$ under a conformal
rescaling $h_{ab}\rightarrow e^{2v}(\psi ^{\ast }\widehat{h})_{ab}$, is
provided by 
\begin{equation}
Z_{(n)}^{\pm }(e^{2v}(\psi ^{\ast }\widehat{h}))=Z_{(n)}^{\pm }(\widehat{h}%
)e^{-2c_{(n)}^{\pm }S_{L}(\widehat{h},v)}  \label{dhoker}
\end{equation}
where 
\begin{equation}
c_{(n)}^{\pm }\doteq 6n^{2}\pm 6n+1
\end{equation}
and where 
\begin{equation}
S_{L}(\widehat{h},v)\doteq \frac{1}{12\pi }\int_{M}d^{2}x\sqrt{\widehat{h}}%
\left[ \frac{1}{2}\widehat{h}^{ik}\partial _{i}v\partial _{k}v+R(\widehat{h}%
)v+\gamma ^{2}(e^{2v}-1)\right] ,
\end{equation}
is the Liouville action (the constant $\gamma $ depends upon the procedure
used for regularizing $Z_{(n)}^{\pm }(h)$). Such a behavior characterizes $%
Z_{(n)}^{\pm }(h)$ as defining on $M$ a conformal field theory with central
charge related to $\pm c_{(n)}^{\pm }$, and is the basic ingredient for
discussing two-dimensional quantum gravity coupled to matter fields with
central charge $c_{mat}$. An analogous though much less transparent
situation holds for the behavior the Weil measure $D_{h}[v]$ under conformal
rescalings $h_{ab}\rightarrow e^{2v}(\psi ^{\ast }\widehat{h})_{ab}$.\ If we
formally assume for $D_{h}[v]$\ transformation properties structurally
similar to (\ref{dhoker}) , (for $n=-1$, and with a Liouville-type term
containing two tunable parameters), then we get the celebrated relation [2],
[3], [4] 
\begin{equation}
Z_{g}[A]=Z_{g}[1]\ A^{\frac{(1-g)}{12}[c_{m}-25-\sqrt{(25-c_{m})(1-c_{m})}%
]-1},  \label{scaling1}
\end{equation}
\ characterizing the scaling, with surface area $A$, of the fixed area
partition function $Z_{g}[A]$ of \ 2D gravity coupled to a matter field of
central charge \ $c_{m}$.\ The quantity \ 
\begin{equation}
\gamma _{string}=\frac{(1-g)}{12}\left( c_{m}-25-\sqrt{(25-c_{m})(1-c_{m})}%
\right) +2,
\end{equation}
\ defines the string susceptibility exponent $\gamma _{string}$ which, for
pure gravity, reduces to the well-known value 
\begin{equation}
\gamma _{string}=\frac{5g-1}{2}.
\end{equation}
The scaling ansatz on $D_{h}[v]$ is believed to be reliable for $c_{m}$
small, a fact reflected in the expression for $\gamma _{string}$ which is
ambiguous for $c_{m}>1$. From the point of view of field theory, we are
entering a region of strong coupling regime between conformal matter and
gravity, and its analysis requires the understanding, not yet achieved, of
the dynamics of Liouville theory. It is an intriguing fact that such
difficulties are absent (or appear in a different, much less problematic
guise) when we approximate the set of Riemannian surfaces $Riem(M)$ with
dynamical triangulations. As a matter of fact, dynamical triangulations (DT)
provide one of the most powerful technique for analyzing two-dimensional
quantum gravity in regimes which are not accessible to the standard
field-theoretic formalism. This is basically due to the circumstance that in
such a discretized setting the quantum measure of the theory, describing the
gravitational dressing of conformal operators in the continuum theory,
reduces to a suitably constrained enumeration of distinct triangulations
admitted by a surface of given topology,(see \emph{e.g.},\ [5] for a
review). And it has been argued, mainly as a consequence of a massive
numerical evidence, that such a (counting) measure automatically accounts
for the anomalous scaling properties of the measure $D\left[ h\right] $ and $%
D_{h}[v]$ governing the continuum path-quantization of 2D gravity. The
geometrical origin of such a property is rather elusive and it is not clear
how the counting for dynamical triangulations factorizes, so to speak, in
terms of a discrete analogous of a moduli space measure and of a Liouville
measure over the conformal degrees of freedom of the theory. Some results in
such a direction have been recently discussed by Catterall and Mottola with
an emphasis on the numerical simulation, [6]. Menotti and Peirano [7]
discussed a similar issue in connection with the Regge measure in simplicial
quantum gravity. However in such a case the problem takes on a rather
different flavor, being more directly connected with the issue of the
diffeomorphism invariance of the resulting measure.

The geometrical explanation of \ the behavior of the DT measure is not
obvious if we only follow the folklore which considers dynamical
triangulations as a sort of \ approximating net in the space of Riemannian
structures $\frac{Riem(M)}{Diff(M)}$. Rather, we \ show that the explanation
is deeply connected with a geometrical mechanism which allow to describe a
dynamically triangulated manifold with $N_{0}$ vertices as a Riemann surface
with $N_{0}$ punctures dressed with a field whose charges describe
discretized curvatures (connected with the deficit angles of the
triangulation). Such a picture calls into play the (compactified) moduli
space of genus $g$ Riemann surfaces with $N_{0}$ punctures $\overline{\mathfrak{M%
}}_{g},_{N_{0}}$, and allow us to prove that the counting of distinct
dynamical triangulations is directly related to the computation of the
Weil-Petersson volume of $\overline{\mathfrak{M}}_{g},_{N_{0}}$. We then exploit
the large $N_{0}$ asymptotics of the Weil-Petersson volume of $\overline{%
\mathfrak{M}}_{g},_{N_{0}}$ recently discussed (in relation with the
Witten-Kontsevich model [8], [9]) by Manin and Zograf \ [10], [11] in order
to prove that the anomalous scaling properties of the counting measure of
dynamical triangulations is only due to the modular degrees of freedom which
parametrizes in $\overline{\mathfrak{M}}_{g},_{N_{0}}$ the vertices of \ the
triangulations. Since these degrees of freedom characterize also the
conformal factor defining the metric geometry of the triangulation, one has
a geometrical explanation of how dynamical triangulations describe the
anomalous scaling of the Weyl measure $D_{h}[v]$. Such an analysis is
discussed here in detail in the case of pure gravity, in the case of
coupling with conformal matter we briefly argue that the role of $\overline{%
\mathfrak{M}}_{g},_{N_{0}}$ is taken over by the moduli space $\overline{\mathfrak{M}%
}_{g},_{N_{0}}(X,\beta )$ of stable maps from the punctured Riemann surface
representing a dynamical triangulation and a (smooth projective) manifold
(variety) $X$ ,(see \emph{e.g.}, [12]). \ Roughly speaking we may think of $%
X $ as the space of possible conformal fields over $M$, and points in $%
\overline{\mathfrak{M}}_{g},_{N_{0}}(X,\beta )$ as representing distributions of
matter fields over dynamical triangulations. According to our analysis in
the pure gravity case, it is rather natural to conjecture that the measure
describing the statistical distribution of such matter fields over the
distinct triangulation is provided by a (generalized) Weil-Petersson volume
of $\overline{\mathfrak{M}}_{g},_{N_{0}}(X,\beta )$. This calls into play the
(descendent)\ Gromov-Witten invariants of $X$ which describe intersection
theory over $\overline{\mathfrak{M}}_{g},_{N_{0}}(X,\beta )$.\ \ This is
interesting from \ various perspectives \ since it is already expected [13]
that $\overline{\mathfrak{M}}_{g},_{N_{0}}(X,\beta )$, via its intersection
theory, is governed by matrix models. Some aspects of the connection between
the anomalous scaling of conformal matter interacting with 2D gravity and
the theory of G-W invariants will be discussed in a forthcoming paper.

\bigskip

\section{\protect\bigskip Triangulations as singular Euclidean structures}

In order to fix the terminology to which we adhere in the rest of the paper,
we need to discuss from a rather unusual perspective the geometry of \
triangulations and polytopal complexes in $2$d simplicial quantum gravity.
Let $T$ denote a $2$-dimensional simplicial complex with underlying
polyhedron $|T|$ and $f$- vector $(N_{0}(T),N_{1}(T),N_{2}(T))$, where $%
N_{i}(T)\in \mathbb{N}$ is the number of $i$-dimensional sub- simplices $%
\sigma ^{i}$ of $T$. A Regge triangulation of a $2$-dimensional PL manifold $%
M$, (without boundary), is a homeomorphism $|T_{l}|\rightarrow {M}\ $ where
each face of $T$ \ \ is geometrically realized by a rectilinear simplex of
variable edge-lengths $l(\sigma ^{1}(k))$\ of the appropriate dimension. A
dynamical triangulation $|T_{l=a}|\rightarrow {M}$ \ is \ a particular case
of a Regge PL-manifold realized by rectilinear and equilateral simplices of
edge-length $l(\sigma ^{1}(k))=$ $a$. The metric structure of a Regge
triangulation\ is locally Euclidean everywhere except at the vertices $%
\sigma ^{0}$, (the \textit{bones}), where the sum of the dihedral angles, $%
\theta (\sigma ^{2})$, of the incident triangles $\sigma ^{2}$'s is in
excess (negative curvature) or in defect (positive curvature) with respect
to the $2\pi $ flatness constraint. The corresponding deficit angle $%
\varepsilon $ is defined by $\varepsilon =2\pi -\sum_{\sigma ^{2}}\theta
(\sigma ^{2})$, where the summation is extended to all $2$-dimensional
simplices incident on the given bone $\sigma ^{0}$. If $K_{T}^{0}$ denotes
the $(0)$-skeleton of $|T_{l}|\rightarrow {M}$, (\emph{i.e.}, the collection
of vertices of the triangulation), then \ $M\backslash {K_{T}^{0}}$ is a
flat Riemannian manifold, and any point in the interior of an $r$- simplex $%
\sigma ^{r}$ has a neighborhood homeomorphic to $B^{r}\times {C}(lk(\sigma
^{r}))$, where $B^{r}$ denotes the ball in $\mathbb{R}^{n}$ and ${C}%
(lk(\sigma ^{r}))$ is the cone over the link $lk(\sigma ^{r})$, (the product 
$lk(\sigma ^{r})\times \lbrack 0,1]$ with $lk(\sigma ^{r})\times \{1\}$
identified to a point), (recall that if we denote by $st(\sigma )$, (the
star of $\sigma $), the union of all simplices of which $\sigma $ is a face,
then $lk(\sigma ^{r})$ is the union of all faces $\sigma ^{f}$\ of the
simplices in $st(\sigma )$ such that $\sigma ^{f}\cap \sigma =\emptyset $).
For dynamical triangulations, the deficit angles are generated by the string
of integers, the \textit{curvature assignments}, $\{q(k)\}_{k=1}^{N_{0}(T)}%
\in \mathbb{N}^{N_{0}(T)}$,\ \textit{viz.},

\begin{equation}
\varepsilon (i)=2\pi -q(i)\arccos (1/2),  \label{curvat}
\end{equation}
where $q(i)\doteq \#\{\sigma ^{2}(h)\bot \sigma ^{0}(i)\}$ provides the
numbers of triangles incident on the $N_{0}(T)$ distinct vertices. For a
regular triangulation \ we have $q(k)\geq 3$, and since each triangle has $3$
vertices $\sigma ^{0}$, the set of integers $\{q(k)\}_{k=1}^{N_{0}(T)}$ \ is
constrained by

\begin{equation}
\sum_{k}^{N_{0}}q(k)=3N_{2}=6\left[ 1-\frac{\chi (M)}{N_{0}(T)}\right]
N_{0}(T),  \label{vincolo}
\end{equation}
where $\chi (M)$ denotes the Euler-Poincar\'{e} characteristic of the
surface, and where $6\left[ 1-\frac{\chi (M)}{N_{0}(T)}\right] $, ($\simeq 6$
for $N_{0}(T)>>1$),\ is the average value of the curvature assignments \ $%
\{q(k)\}_{k=1}^{N_{0}}$.

\begin{remark}
Note that in what follows we shall consider semi-simplicial complexes for
which the constraint $q(k)\geq 3$ is removed. Examples of such
configurations are afforded by triangulations with pockets, where two
triangles are incident on a vertex, or more generally by triangulations
where the star of a vertex may contain just one triangle. We shall refer to
such extended configurations as generalized (Regge and dynamical)
triangulations\ .
\end{remark}

In this connection, it is well known that in dimension two regular
triangulations and generalized triangulations simply provide different
approximations of \ the same continuum quantum gravity theory [5]. Thus it
would seem natural to restrict attention to the simpler ensemble of \ pure
simplicial complexes. Such a restriction is however quite artificial, and in
discussing the geometrical aspects of \ simplicial quantum gravity (for
instance its connection with moduli space theory [8], [9]),\ the set of
semi-simplicial complexes provide a more natural setup for our analysis.

\bigskip

\ As recalled, a (generalized) Regge triangulation $|T_{l}|\rightarrow M$
defines on the PL manifold $M$ a polyhedral metric with conical
singularities, associated with the vertices $\{\sigma
^{0}(i)\}_{i=1}^{N_{0}(T)}$ of the triangulation, but which is otherwise
flat and smooth everywhere else. Such a metric has important special
features, in particular it induces on the PL manifold \ $M$ a geometrical
structure which turns out to be a particular case of the theory of \emph{%
Singular \ Euclidean Structure} (in the sense of M. Troyanov [14], and W.
Thurston [15]).\ This structure can be most conveniently described in terms
of complex function theory. To this end, let us consider the (first)
barycentric subdivision of \ $T$, then the closed stars, in such a
subdivision, of the vertices of the original triangulation $T_{l}$ form a
collection of $2$-cells $\{\rho ^{2}(i)\}_{i=1}^{N_{0}(T)}$ characterizing\
a polytope $P$ barycentrically dual to $T$.

\begin{remark}
Note that here we are not considering a rectilinear presentation of the dual
cell complex $P$ (where the PL-polytope is realized by flat polygonal $2$%
-cells $\{\rho ^{2}(i)\}_{i=1}^{N_{0}(T)}$) but rather a geometrical
presentation $|P_{T_{L}}|\rightarrow {M}$ of $P$ where the $2$-cells $\{\rho
^{2}(i)\}_{i=1}^{N_{0}(T)}$ retain the conical geometry induced on the
barycentric subdivision by the original metric structure of $%
|T_{l}|\rightarrow {M}$. Namely, if $(\lambda (k),\chi (k))$ denote polar
coordinates (based at $\sigma ^{0}(k)$) of $p\in \rho ^{2}(k)$, then $\rho
^{2}(k)$ is geometrically realized as the space 
\begin{equation}
\frac{\left\{ (\lambda (k),\chi (k))\ :\lambda (k)\geq 0;\chi (k)\in 
\mathbb{R}/(2\pi -\varepsilon (k))\mathbb{Z}\right\}}{(0,\chi
(k))}\sim (0,\chi ^{\prime }(k))
\end{equation}
endowed with the metric 
\begin{equation}
d\lambda (k)^{2}+\lambda (k)^{2}d\chi (k)^{2}.
\end{equation}
This definition characterizes the \emph{conical} Regge polytope $%
|T_{l}|\rightarrow {M}$ (and its rigid equilateral specialization $%
|P_{T_{a}}|\rightarrow {M}$) \ barycentrically dual to $|T_{l}|\rightarrow {M%
}$.
\end{remark}

In order to switch back and forth between the geometry of $%
|T_{l}|\rightarrow {M}$ and the corresponding conical geometry of the
barycentrically dual polytope $|P_{T_{L}}|\rightarrow {M}$ it is sufficient
to relate the lengths of the sides of the generic triangle $\sigma
^{2}(\alpha )\in |T_{l}|\rightarrow {M}$ to the lengths of the corresponding
medians connecting the barycenter $\rho ^{0}(\alpha )$ of $\sigma
^{2}(\alpha )$ to the barycenters of the sides of $\sigma ^{2}(\alpha )$. If
we denote by $l_{i}(\alpha )$, $i=1,2,3$, the lengths of such sides, and by $%
\widehat{L}_{i}(a)$ the lengths of the corresponding medians, (\emph{e.g.}, $%
\widehat{L}_{1}(a)$ is the length of the median connecting $\rho ^{0}(\alpha
)$ with the barycenter of $l_{1}(\alpha )$), then a direct computation
provides 
\begin{equation}
\begin{tabular}{ccc}
$\widehat{L}_{1}^{2}(a)$ & $=$ & $\frac{1}{18}l_{3}^{2}(\alpha )+\frac{1}{18}%
l_{2}^{2}(\alpha )-\frac{1}{36}l_{1}^{2}(\alpha )$ \\ 
$\widehat{L}_{2}^{2}(a)$ & $=$ & $\frac{1}{18}l_{1}^{2}(\alpha )+\frac{1}{18}%
l_{3}^{2}(\alpha )-\frac{1}{36}l_{2}^{2}(\alpha )$ \\ 
$\widehat{L}_{3}^{2}(a)$ & $=$ & $\frac{1}{18}l_{1}^{2}(\alpha )+\frac{1}{18}%
l_{2}^{2}(\alpha )-\frac{1}{36}l_{3}^{2}(\alpha )$ \\ 
&  &  \\ 
$l_{1}^{2}(\alpha )$ & $=$ & $8\widehat{L}_{3}^{2}(a)+8\widehat{L}%
_{2}^{2}(a)-4\widehat{L}_{1}^{2}(a)$ \\ 
$l_{2}^{2}(\alpha )$ & $=$ & $8\widehat{L}_{1}^{2}(a)+8\widehat{L}%
_{3}^{2}(a)-4\widehat{L}_{2}^{2}(a)$ \\ 
$l_{3}^{2}(\alpha )$ & $=$ & $8\widehat{L}_{1}^{2}(a)+8\widehat{L}%
_{2}^{2}(a)-4\widehat{L}_{3}^{2}(a)$%
\end{tabular}
.
\end{equation}

In particular, if we denote by $L(\alpha ,\beta )$ the length of the edge $%
\rho ^{1}(\alpha ,\beta )$\break $\in |P_{T_{L}}|\rightarrow {M}$ connecting the
barycenters $\rho ^{0}(\alpha )$ and $\rho ^{0}(\beta )$ of two triangles $%
\sigma ^{2}(\alpha )$ and $\sigma ^{2}(\beta )$ in $|T_{l}|\rightarrow {M}$
sharing the sides $l_{3}(\alpha )$ and $l_{1}(\beta )$, then $L(\alpha
,\beta )=$ $\widehat{L}_{3}(a)+\widehat{L}_{1}(\beta )$, and we get 
\begin{equation}
L(\alpha ,\beta )=\frac{1}{6}\left( \sqrt{2l_{1}^{2}(\alpha
)+2l_{2}^{2}(\alpha )-l_{3}^{2}(\alpha )}+\sqrt{2l_{3}^{2}(\beta
)+2l_{2}^{2}(\beta )-l_{1}^{2}(\beta )}\right) .
\end{equation}

In such a geometrical setup, let $\rho ^{2}(k)$ be the generic two-cell $\in
|P_{T_{L}}|\rightarrow {M}$ barycentrically dual to the vertex $\sigma
^{0}(k)\in |T_{l}|\rightarrow M$ and let us denote by 
\begin{equation}
L(\partial (\rho ^{2}(k)))=\sum_{h=1}^{q(k)}L(\rho ^{1}(h))
\end{equation}
\ the length of the boundary $\partial (\rho ^{2}(k))$\ of $\rho ^{2}(k)$,
where $L(\rho ^{1}(h))$ are the lengths of the $q(k)$ ordered edges $\{\rho
^{1}(j)\}$ $\in \partial (\rho ^{2}(k))$. \ If $\varepsilon (k)$ denotes the
deficit angle corresponding to $\sigma ^{0}(k)\in |T_{l}|\rightarrow M$,
then we define the slant radius associated with the cell $\rho ^{2}(k)\in
|P_{T_{L}}|\rightarrow {M}$ according to 
\begin{equation}
r(k)\doteq \frac{L(\partial (\rho ^{2}(k)))}{2\pi -\varepsilon (k)}.
\label{efradius}
\end{equation}
\ \ Let 
\begin{equation}
B^{2}(k)\doteq \left\{ p\in \rho ^{2}(k)\backslash \partial (\rho
^{2}(k))\right\}  \label{ball}
\end{equation}
\ \ the open ball (a Euclidean cone) associated with $\rho ^{2}(k)$ and
contained in the star $st(\sigma ^{0}(k))$ of the given vertex $\sigma
^{0}(k)$.\ Note that any two such balls, say $B^{2}(k)$ and $B^{2}(j)$, $%
k\neq j$, are pairwise disjoint, and that the complex 
\begin{equation}
\left( |T_{l}|\rightarrow M\right) /\bigcup_{k=1}^{N_{0}(T)}B^{2}(k)
\end{equation}
\ retracts on the $1$-skeleton $K^{1}\left[ P_{T_{L}}\right] $ of $%
|P_{T_{L}}|\rightarrow {M}$. To any vertex $\sigma ^{0}(k)\in $\ $%
|T_{l}|\rightarrow M$ \ we associate a complex uniformizing coordinate $%
t_{k}\in \mathbb{C}$ defined in the open disk of radius $r(k)$, \emph{viz}. 
\begin{gather}
B^{2}(k)\overset{t_{k}}{\longrightarrow }D_{k}(r(k))\doteq \left\{ t_{k}\in 
\mathbb{C}\;|0\leq t_{k}<r(k)\right\} \\
B^{2}(k)\ni p\longmapsto t_{k}(p).  \notag
\end{gather}
In terms of \ $t_{k}$ we can explicitly write down the singular Euclidean
metric locally characterizing the singular Euclidean structure of $\
B^{2}(k) $, according to 
\begin{equation}
ds_{(k)}^{2}\doteq e^{2u}\left| t_{k}-t_{k}(\sigma ^{0}(k))\right|
^{-2\left( \frac{\varepsilon (k)}{2\pi }\right) }\left| dt_{k}\right| ^{2},
\label{cmetr2}
\end{equation}
where \ $\varepsilon (k)$ is given by (\ref{curvat}), and $%
u:B^{2}\rightarrow \mathbb{R}$ is a continuous function ($C^{2}$ on $%
B^{2}-\{\sigma ^{0}(k)\}$) such that, for $t_{k}\rightarrow t_{k}(\sigma
^{0}(k))$, 
\begin{gather}
\left| t_{k}-t_{k}(\sigma ^{0}(k))\right| \frac{\partial u}{\partial t_{k}}%
\rightarrow 0, \\
\left| t_{k}-t_{k}(\sigma ^{0}(k))\right| \frac{\partial u}{\partial 
\overline{t_{k}}}\rightarrow 0.  \notag
\end{gather}
Up to the presence of the normalizing conformal factor $e^{2u}$, one
recognizes in (\ref{cmetr2}) the metric of a Euclidean cone of total angle $%
\theta (k)=2\pi -\varepsilon (k)$. We can glue together the uniformizations $%
\{D_{k}(r(k))\}_{k=1}^{N_{0}(T)}$ along the pattern defined by the $1$%
-skeleton of $|P_{T_{L}}|\rightarrow {M}$ and generate on $M$ the
quasi-conformal structure 
\begin{equation}
\left( M,\mathcal{C}_{sg}\right) \underset{|P_{T_{L}}|\rightarrow {M}}{%
\doteq \bigcup }\{D_{k}(r(k));ds_{(k)}^{2}\}_{k=1}^{N_{0}(T)}  \label{sing}
\end{equation}
naturally associated with $|T_{l}|\rightarrow M$. \ If $|dt|^{2}$ is a
smooth (conformally flat) metric on $M$, then $\left( M,\mathcal{C}%
_{sg}\right) $\ \ can be (locally) represented by the metric 
\begin{equation}
ds_{T}^{2}=e^{2v}|dt|^{2},  \label{cmetr}
\end{equation}
where the conformal factor $v$ is given by\ 
\begin{equation}
v\doteq u-\sum_{k=1}^{N_{0}(T)}\left( \frac{\varepsilon (k)}{2\pi }\right)
\ln \left| t-t_{k}\right| .  \label{conform}
\end{equation}
Even if $\left( M,\mathcal{C}_{sg}\right) $ is uniquely defined by such
metric, the singular metric $e^{2v}|dt|^{2}$ characterizing a given $\left(
M,\mathcal{C}_{sg}\right) $\ is only defined up to the conformal symmetry of 
$\left( M,\mathcal{C}_{sg}\right) $. For instance for $M\approx \mathbb{S}%
^{2}$ we have a residual $SL(2,\mathbb{C})$ invariance, [7]. \ It is easily
verified that the number of degrees of freedom associated with (\ref{cmetr})
corresponds indeed to the number $N_{1}(T)$\ of edges of the Regge
triangulation $|T_{l=a}|\rightarrow M$. The actual count follows by
observing that \ (\ref{cmetr}) is parametrized by the $N_{0}(T)$ complex
coordinates \ $t_{k}$ of the conical singularities, by the $N_{0}(T)$
deficit angles $\varepsilon (k)$ constrained by $\sum_{k=1}^{N_{0}(T)}\left(
-\frac{\varepsilon (k)}{2\pi }\right) =2g-2$, by the conformal factor $u$
(see (\ref{conform})), and by the $6g-6$ moduli which parametrize the
possible inequivalent choices of the smooth base metric $|dt|^{2}$ in \ (\ref
{cmetr}), (assume that $g\geq 2$, otherwise in such a count we have also to
take into account the dimension of the space of conformal Killing vector
fields of $|dt|^{2}$). This sums up to 
\begin{equation}
3N_{0}(T)+6g-6,
\end{equation}
\ \ degrees of freedom which, according to the Dehn-Sommerville relations $\
N_{0}(T)-N_{1}(T)+N_{2}(T)=2-2g$, and $2N_{1}(T)=3N_{2}(T)$, actually equals 
$N_{1}(T)$ .\ 

\ 

The singular structure of the metric defined by (\ref{cmetr}) and (\ref
{conform}) can be naturally summarized in a formal linear combination of the
points $\{\sigma ^{0}(k)\}$ with coefficients given by the corresponding
deficit angles (normalized to $2\pi $), \emph{viz.}, in\ the \emph{real
divisor }[14] 
\begin{equation}
Div(T)\doteq \sum_{k=1}^{N_{0}(T)}\left( -\frac{\varepsilon (k)}{2\pi }%
\right) \sigma ^{0}(k)=\sum_{k=1}^{N_{0}(T)}\left( \frac{\theta (k)}{2\pi }%
-1\right) \sigma ^{0}(k)
\end{equation}
supported on the set of bones $\{\sigma ^{0}(i)\}_{i=1}^{N_{0}(T)}$. Note
that the degree of such a divisor, \ defined by 
\begin{equation}
\left| Div(T)\right| \doteq \sum_{k=1}^{N_{0}(T)}\left( \frac{\theta (k)}{%
2\pi }-1\right) =-\chi (M)  \notag
\end{equation}
is, for dynamical triangulations, a rewriting of the combinatorial
constraint (\ref{vincolo}). In such a sense, the pair $(|T_{l}|\rightarrow
M,Div(T))$, or shortly, $(T,Div(T))$, encodes the datum of the triangulation 
$|T_{l}|\rightarrow M$ and of a corresponding set of curvature assignments $%
\{q(k)\}$ on the bones $\{\sigma ^{0}(i)\}_{i=1}^{N_{0}(T)}$. \ The real
divisor $\left| Div(T)\right| $ characterizes the Euler class of \ the pair\ 
$(T,Div(T))$ and yields for a corresponding Gauss-Bonnet formula.
Explicitly, the Euler number associated with $(T,Div(T))$ is defined, [14],
by

\begin{equation}
e(T,Div(T))\doteq \chi (M)+|Div(T)\mathbf{|.}  \label{euler}
\end{equation}
and the Gauss-Bonnet formula reads [14]:

\begin{lemma}
(\textbf{Gauss-Bonnet for triangulated surfaces}) 
\\ Let $(T,Div(T))$ be a triangulated surface with divisor 
\begin{equation}
Div(T)\doteq \sum_{k=1}^{N_{0}(T)}\left( \frac{\theta (k)}{2\pi }-1\right)
\sigma ^{0}(k),
\end{equation}
associated with the vertices incidences $\{\sigma ^{0}(k)\}_{k=1}^{N_{0}(T)}$%
. Let $ds^{2}$ be the conformal metric (\ref{cmetr}) representing the
divisor $Div(T)$ . Then 
\begin{equation}
\frac{1}{2\pi }\int_{M}KdA=e(T,Div(T)),
\end{equation}
where $K$\ and $dA$\ respectively are the curvature and the area element
corresponding to the metric $ds_{T}^{2}.$
\end{lemma}

Note that such a theorem holds for any singular Riemann surface $\Sigma $
described by a divisor $Div(\Sigma )$ and not just for triangulated surfaces
[14]. Since for a Regge (dynamical) triangulation, we have $%
e(T_{a},Div(T))=0 $, the Gauss-Bonnet formula implies

\begin{equation}
\frac{1}{2\pi }\int_{M}KdA=0.
\end{equation}
Thus, a triangulation $|T_{l}|\rightarrow M$ naturally carries a conformally
flat structure. Clearly this is a rather obvious result, (since the metric
in $M-\{\sigma ^{0}(i)\}_{i=1}^{N_{0}(T)}$ is flat). However, it admits a
not-trivial converse (recently proved by M. Troyanov, but, in a sense, going
back to E. Picard) [14], [16]:

\begin{theorem}
(\textbf{Troyanov-Picard}) Let $(\left( M,\mathcal{C}_{sg}\right) ,Div)$ be
a singular Riemann surface with a divisor such that $e(M,Div)=0$. Then there
exists on $M$\ a unique (up to homothety) conformally flat metric
representing the divisor $Div$.
\end{theorem}

\noindent These results geometrically characterize metrical triangulations (and the
associated conical Regge polytopal surfaces) as a particular case of the
theory of singular Riemann surfaces, and provides the rationale for viewing
two-dimensional simplicial quantum gravity in a more analytic spirit. In
order to put this latter remark in a proper perspective, let us set 
\begin{equation}
(M;N_{0})\doteq M-\{\sigma ^{0}(i)\}_{i=1}^{N_{0}(T)},
\end{equation}
with $2-2g-N_{0}(T)<0$. Also, let us denote by $\mathcal{H}^{2}=\{\zeta \in 
\mathbb{C\;}|\func{Im}(\zeta )>0\}$ the upper half-plane equipped with the
metric $h(\zeta )\left| d\zeta \right| ^{2}\doteq \frac{|d\zeta |^{2}}{(%
\func{Im}(\zeta ))^{2}}$. \ As is well known, the set of fractional linear
(M\"{o}bius) transformations $PSL(2,\mathbb{R})=SL(2,\mathbb{R})/\{\pm I\}$
acts on $(\ \mathcal{H}^{2},h(\zeta )\left| d\zeta \right| ^{2})$ by
homographic transformations and coincides with the group of orientation
preserving isometries of the hyperbolic upper half-plane\ $\ (\ \mathcal{H}%
^{2},h(\zeta )\left| d\zeta \right| ^{2})$. According to the
Poincar\'{e}-Klein-Koebe uniformization theorem\ any Riemann surface $M$ (of
genus $g\geq 2$) can be represented as a quotient $\mathcal{H}^{2}/\Gamma $,
where $\Gamma $ is a discrete subgroup of $PSL(2,\mathbb{R})$ acting freely
on $(\ \mathcal{H}^{2},h(\zeta )\left| d\zeta \right| ^{2})$, and the metric 
$h(\zeta )\left| d\zeta \right| ^{2}$ descends to the quotient defining a
complete metric of constant curvature $-1$ on $\mathcal{H}^{2}/\Gamma $. \
The discrete subgroup $\Gamma $ is canonically isomorphic to the fundamental
group $\pi _{1}(M)$ of the surface $M=\mathcal{H}^{2}/\Gamma $, and any
homotopy class of closed curves on $M=(\mathcal{H}^{2}/\Gamma ,h(\zeta
)\left| d\zeta \right| ^{2})$ contains a unique geodesic. Any two surfaces $(%
\mathcal{H}^{2}/\Gamma )$ and $(\mathcal{H}^{2}/\Gamma ^{\prime })$
resulting from such a construction are isomorphic if and only if the
subgroups $\Gamma $ and $\Gamma ^{\prime }$ are conjugated by an element of $%
PSL(2,\mathbb{R})$. Let us denote by $\mathfrak{T}_{g}(M)$ the Teichm\"{u}ller
space of all conformal structures on $M$ under the equivalence relation
given by pullback by diffeomorphisms isotopic to the identity map $%
id:M\rightarrow M$. The standard uniformization \ $M=(\mathcal{H}^{2}/\Gamma
,h(\zeta )\left| d\zeta \right| ^{2})$ provides an embedding of $\mathfrak{T}%
_{g}(M)$ into a connected component of the representation variety 
\begin{equation}
\frac{Hom(\pi _{1}(M),PSL(2,\mathbb{R}))}{PSL(2,\mathbb{R})},  \label{repvar}
\end{equation}
the set of conjugacy classes of homomorphisms 
$$\Phi :\pi _{1}(M)\rightarrow PSL(2,\mathbb{R}).$$ 
The component characterized by $\mathfrak{T%
}_{g}(M)$ is the one associated with conjugacy classes of representations $%
[\Phi ]$ for which $\Phi (\lambda )$ is an hyperbolic element of $PSL(2,%
\mathbb{R})$, (\emph{i.e.}, such that the trace of the corresponding matrix
is $>2$), whenever $\lambda $ is a non trivial homotopy class in $\pi
_{1}(M) $. As stressed by W. Goldman, [17] the remaining components of the
representation variety (\ref{repvar}) are related to the theory of singular
geometric structures developed by Troyanov and Thurston, and briefly
described above. In particular, each component in (\ref{repvar}) is
classified by a corresponding Euler class (\ref{euler}). As we have seen,
for dynamical (and Regge) triangulations $|T_{l}|\rightarrow M$ the Euler
class is zero, and there exists on $(M;N_{0})\ $a unique non-singular
Euclidean structure. The corresponding homomorphism 
\begin{equation}
\pi _{1}\left( (M;N_{0})\right) \longrightarrow PSL(2,\mathbb{R})
\label{holon}
\end{equation}
is defined by sending the link of $\sigma ^{0}(i)$ into the elliptic element
of $PSL(2,\mathbb{R})$ which describes the rotation of angle $\theta (i)$
around $\sigma ^{0}(i)$, (recall that a matrix of $PSL(2,\mathbb{R})$ is
elliptic if and only if it is conjugate in $SL(2,\mathbb{R})$ to a unique
rotation matrix; note also that for any $x\in \mathcal{H}^{2}$, the
stabilizer group of $x$ in $PSL(2,\mathbb{R})$ is conjugate to the circle
group $SO(2)$). Thus, by means of the holonomy map (\ref{holon}), the
singular Riemann surface $(\left( M,\mathcal{C}_{sg}\right) ,Div)$\ \
describes a (generalized) triangulation $|T_{l}|\rightarrow M$ \ as a
singular uniformization of the Euclidean surface $(M;N_{0})$. \ The
advantage of this approach is that it directly relates simplicial quantum
gravity to the properties of the Teichm\"{u}ller space associated with the
Riemann surfaces $(\left( M,\mathcal{C}_{sg}\right) ,Div)$ describing the
inequivalent triangulations $|T_{l}|\rightarrow M$. This is quite appealing
since brings simplicial quantum gravity even closer to continuum $2$D
gravity, where Riemann surface theory plays a prominent role.\ The
disadvantage is that here we are not dealing with ordinary Teichm\"{u}ller
space theory, since the relevant component of \ (\ref{repvar}) refers to
Euclidean structures with conical singularities. As recalled, these latter
are in the connected component of the representation variety 
\begin{equation}
\frac{Hom(\pi _{1}((M;N_{0})),PSL(2,\mathbb{R}))}{PSL(2,\mathbb{R})},
\label{variety}
\end{equation}
with Euler class $0$, and we need non-standard techniques from the theory of
geometric structures in order to work in such a setting. A typical example
in this direction is provided by Thurston analysis of the space of
dynamically triangulated spheres with positive deficit angles [15]. In this
connection it is also interesting to remark that moduli spaces of singular
Euclidean structures (again on the $2$-sphere) occurs in the classical
analysis of the monodromy properties of hypergeometric functions carried out
by Deligne and Mostow [18].

There is a way of \ circumventing the use of sophisticated
representation theoretic techniques by relating the space of inequivalent
singular Euclidean structures to the more standard theory of punctured
Riemann surfaces. This boils down to the observation that on punctured
Riemann surfaces we can introduce flat metrics by exploiting the connection
between ribbon graphs and the theory of Jenkins-Strebel (JS) quadratic
differentials [19], [20]. Such a correspondence is well-known in $2$-D
quantum gravity where it plays a crucial role in Kontsevich's proof [8] of
the Witten conjecture; see Loijenga [21] for details. In the spirit of our
paper, we follow a slightly different approach emphasizing more the analytic
aspects of the theory.

\subsection{Conical Regge polytopes and ribbon graphs}

\ Let us start by recalling that the geometrical realization of the $1$%
-skeleton of \ the conical Regge polytope\ $|P_{T_{L}}|\rightarrow {M}$ is a 
$3$-valent graph 
\begin{equation}
\Gamma =(\{\rho ^{0}(k)\},\{\rho ^{1}(j)\})
\end{equation}
where the vertex set $\{\rho ^{0}(k)\}_{k=1}^{N_{2}(T)}$ is identified with
the barycenters of the triangles $\{\sigma ^{o}(k)\}_{k=1}^{N_{2}(T)}\in
|T_{l}|\rightarrow M$, whereas each edge $\rho ^{1}(j)\in \{\rho
^{1}(j)\}_{j=1}^{N_{1}(T)}$ is generated by two half-edges $\rho ^{1}(j)^{+}$
and $\rho ^{1}(j)^{-}$ joined through the barycenters $\{W(h)%
\}_{h=1}^{N_{1}(T)}$ of the edges $\{\sigma ^{1}(h)\}$ belonging to the
original triangulation $|T_{l}|\rightarrow M$. Thus, if we formally
introduce a degree-$2$ vertex at each middle point $\{W(h)%
\}_{h=1}^{N_{1}(T)} $, the actual graph naturally associated to the $1$%
-skeleton of \ $|P_{T_{L}}|\rightarrow {M}$ is 
\begin{equation}
\Gamma _{ref}=\left( \{\rho
^{0}(k)\}\bigsqcup_{h=1}^{N_{1}(T)}\{W(h)\},\{\rho
^{1}(j)^{+}\}\bigsqcup_{j=1}^{N_{1}(T)}\{\rho ^{1}(j)^{-}\}\right) ,
\end{equation}
the so called edge-refinement [20] of $\Gamma =(\{\rho ^{0}(k)\},\{\rho
^{1}(j)\})$. The relevance of such a notion stems from the observation that
the natural automorphism group $Aut(P_{L})$ of \ $|P_{T_{L}}|\rightarrow {M}$%
, (\emph{i.e.}, the set of bijective maps $\Gamma =(\{\rho ^{0}(k)\},\{\rho
^{1}(j)\})\rightarrow \widetilde{\Gamma }=(\widetilde{\{\rho ^{0}(k)\}},%
\widetilde{\{\rho ^{1}(j)\}}$ preserving the incidence relations defining
the graph structure), is not the automorphism group of $\Gamma $ but rather
the (larger) automorphism group of its edge refinement [20], \emph{i.e.}, 
\begin{equation}
Aut(P_{L})\doteq Aut(\Gamma _{ref}).
\end{equation}
The locally uniformizing complex coordinate $t_{k}\in \mathbb{C}$ \ in terms
of which we can explicitly write down the singular Euclidean metric (\ref
{cmetr}) around each vertex $\sigma ^{0}(k)\in $ $|T_{l}|\rightarrow M$,
provides a (counterclockwise) orientation in the $2$-cells of $%
|P_{T_{L}}|\rightarrow {M}$. Such an orientation gives rise to a cyclic
ordering on the set of half-edges $\{\rho ^{1}(j)^{\pm }\}_{j=1}^{N_{1}(T)}$
incident on the vertices $\{\rho ^{0}(k)\}_{k=1}^{N_{2}(T)}$. \ According to
these remarks, the $1$-skeleton of \ $|P_{T_{L}}|\rightarrow {M}$ is a
ribbon (or fat) graph [5], \emph{viz.}, a graph $\Gamma $ together with a
cyclic ordering on the set of half-edges incident to each vertex of $\
\Gamma $. Conversely, any ribbon graph $\Gamma $ characterizes an oriented
surface $M(\Gamma )$ with boundary possessing $\Gamma $ as a spine, (\emph{%
i.e.}, the inclusion $\Gamma \hookrightarrow M(\Gamma )$ is a homotopy
equivalence). This is an appropriate place to note that not all trivalent
metric ribbon graphs are barycentrically dual to regular triangulations. We
may have trivalent metric ribbon graphs with digons and loops. From the
point of view of the theory of singular Euclidean structures, (which is the
real geometrical category underlying the use of simplicial methods in
gravity), there is no obvious reason to get rid, a priori, of such more
general configurations. As already mentioned, this is the reason for which
we must extend our analysis to generalized triangulations, (see remark 1),
and to their associated barycentrically dual conical Regge polytopes . In
this way (the edge-refinement of) the $1$-skeleton of a generalized conical
Regge polytope\ $|P_{T_{L}}|\rightarrow {M}$ is in a one-to-one
correspondence with trivalent metric ribbon graphs.

We can associate with $|P_{T_{L}}|\rightarrow {M}$ a complex structure $%
((M;N_{0}),\mathcal{C})$ (a punctured Riemann surface)\ which is, in a
well-defined sense, dual to the structure $\left( M,\mathcal{C}_{sg}\right) $
generated by $|T_{l}|\rightarrow M$. Note that according to our
characterization of \ $|P_{T_{L}}|\rightarrow {M}$, the $2$-cells $\{\rho
^{2}(k)\}$\ have the conical geometry (\ref{cmetr2}), and the strategy for
defining $((M;N_{0}),\mathcal{C})$ is to desingularize $\left( M,\mathcal{C}%
_{sg}\right) $ by transforming the conical singularities into a suitable
decoration of $((M;N_{0}),\mathcal{C})$. In order to see in detail such a
construction, let $\rho ^{2}(k)$ be the generic two-cell $\in
|P_{T_{L}}|\rightarrow {M}$ barycentrically dual to the vertex $\sigma
^{0}(k)\in |T_{l}|\rightarrow M$ .\ To the generic edge $\rho ^{1}(h)$ of $%
\rho ^{2}(k)$\ \ we associate a complex uniformizing coordinate $z(h)$
defined in the strip 
\begin{equation}
U_{\rho ^{1}(h)}\doteq \{z(h)\in \mathbb{C}|0<\func{Re}z(h)<L(\rho
^{1}(h))\},
\end{equation}
$L(\rho ^{1}(h))$ being the length of the edge considered. The uniformizing
coordinate $w(j)$, corresponding to the generic $3$-valent vertex $\rho
^{0}(j)\in \rho ^{2}(k)$,\ \ is defined in the open set 
\begin{equation}
U_{\rho ^{0}(j)}\doteq \{w(j)\in \mathbb{C}|\;|w(j)|<\delta ,\;w(j)[\rho
^{0}(j)]=0\},
\end{equation}
where $\delta >0$ is a suitably small constant. Finally, the two-cell $\rho
^{2}(k)$\ is uniformized in the unit disk 
\begin{equation}
U_{\rho ^{2}(k)}\doteq \{\zeta (k)\in \mathbb{C}|\;|\zeta (k)|<1,\;\zeta
(k)[\sigma ^{0}(k)]=0\},
\end{equation}
where $\sigma ^{0}(k)$ is the vertex $\in |T_{l}|\rightarrow M$ \
corresponding to the given two-cell.In order to coherently glue together 
$$\{w(j),U_{\rho ^{0}(j)}\}_{j=1}^{N_{2}(T)},\{z(h),U_{\rho
^{1}(h)}\}_{h=1}^{N_{1}(T)} \;\textrm{and}\; \{\zeta (k),U_{\rho
^{2}(k)}\}_{k=1}^{N_{0}(T)}$$
one exploits the connection between ribbon
graphs and quadratic differentials. An appropriate way to proceed is to note
that to each edge $\rho ^{1}(h)\in $ $\rho ^{2}(k)$\ we can associate the
standard quadratic differential on $U_{\rho ^{1}(h)}$ given by 
\begin{equation}
\psi (h)|_{\rho ^{1}(h)}=dz(h)\otimes dz(h).  \label{foliat}
\end{equation}
Such $\psi (h)|_{\rho ^{1}(h)}$ can be extended to the remaining local
uniformizations $U_{\rho ^{0}(j)}$, and $U_{\rho ^{2}(k)}$,\ \ by exploiting
a classic result in Riemann surface theory according to which a quadratic
differential $\psi $ has a finite number of zeros $n_{zeros}(\psi )$\ with
orders\ $k_{i}$ and a finite number of poles $n_{poles}(\psi )$ of order $%
s_{i}$ such that 
\begin{equation}
\sum_{i=1}^{n_{zero}(\psi )}k_{i}-\sum_{i=1}^{n_{pole}(\psi )}s_{i}=4g-4.
\label{quadrel}
\end{equation}
\ In our case we must have $n_{zeros}(\psi )=N_{2}(T)$ with $k_{i}=1$,
(corresponding to the fact that the $1$-skeleton of $|P_{L}|\rightarrow M$
is a trivalent graph), and \ $n_{poles}(\psi )=$\ $N_{0}(T)$ with $%
s_{i}=s\forall i$, for a suitable positive integer $s$. According to such
remarks (\ref{quadrel}) reduces to 
\begin{equation}
N_{2}(T)-sN_{0}(T)=4g-4.  \label{poles}
\end{equation}
However, from the Euler relation $N_{0}(T)-N_{1}(T)+N_{2}(T)=2-2g$, and $%
2N_{1}(T)=3N_{2}(T)$ we get $N_{2}(T)-2N_{0}(T)=4g-4$. \ This is consistent
with (\ref{poles}) if and only if\ \ $s=2$. Thus the extension $\psi $ of $%
\psi (h)|_{\rho ^{1}(h)}$ along the $1$-skeleton of $|P_{L}|\rightarrow M$
must have $N_{2}(T)$ zeros of order $1$ corresponding to the trivalent
vertices $\{\rho ^{0}(j)\}$\ of $|P_{L}|\rightarrow M$ and $N_{0}(T)$
quadratic poles corresponding to the polygonal cells $\{\rho ^{2}(k)\}$ of
perimeter lengths $\{L(\partial (\rho ^{2}(k)))\}$. Around a zero of order
one and a pole of order two, every quadratic differential $\psi $ has a
canonical local structure which (along with (\ref{foliat})) is summarized in
the following table [20] \ 
\begin{equation}
(|P_{T_{L}}|\rightarrow {M)\rightarrow }\psi \doteq \left\{ 
\begin{tabular}{l}
$\psi (h)|_{\rho ^{1}(h)}=dz(h)\otimes dz(h),$ \\ 
$\psi (j)|_{\rho ^{0}(j)}=\frac{9}{4}w(j)dw(j)\otimes dw(j),$ \\ 
$\psi (k)|_{\rho ^{2}(k)}=-\frac{\left[ L(\partial (\rho ^{2}(k)))\right]
^{2}}{4\pi ^{2}\zeta ^{2}(k)}d\zeta (k)\otimes d\zeta (k),$%
\end{tabular}
\right.   \label{differ}
\end{equation}
where $\{\rho ^{0}(j),\rho ^{1}(h),\rho ^{2}(k)\}$ runs over the set of
vertices, edges, and $2$-cells of $|P_{L}|\rightarrow M$. Since $\psi
(h)|_{\rho ^{1}(h)}$, $\psi (j)|_{\rho ^{0}(j)}$, and $\psi (k)|_{\rho
^{2}(k)}$ must be identified on the non-empty pairwise intersections $%
U_{\rho ^{0}(j)}\cap U_{\rho ^{1}(h)}$, $U_{\rho ^{1}(h)}\cap U_{\rho
^{2}(k)}$ we can associate to the polytope $|P_{T_{L}}|\rightarrow {M}$\ a
complex structure $(M;N_{0},\mathcal{C})$\ by coherently glueing, along the
pattern associated with the ribbon graph $\Gamma $, the local
uniformizations $\{U_{\rho ^{0}(j)}\}_{j=1}^{N_{2}(T)}$, $\{U_{\rho
^{1}(h)}\}_{h=1}^{N_{1}(T)}$, and \ $\{U_{\rho ^{2}(k)}\}_{k=1}^{N_{0}(T)}$.
Explicitly,\ let $\{U_{\rho ^{1}(h_{\alpha })}\}$, $\alpha =1,2,3$ be the
three generic open strips associated with the three cyclically oriented
edges $\{\rho ^{1}(h_{\alpha })\}$ incident on the generic vertex $\rho
^{0}(j)$. Then the uniformizing coordinates $\{z(h_{\alpha })\}$ are related
to $w(j)$ by the transition functions 
\begin{equation}
w(j)=e^{2\pi i\frac{\alpha -1}{3}}z(h_{\alpha })^{\frac{2}{3}},\hspace{0.2in}%
\hspace{0.1in}\alpha =1,2,3.  \label{glue1}
\end{equation}
Note that in such uniformization the vertices $\{$ $\rho ^{0}(j)\}$ do not
support conical singularities since each strip $U_{\rho ^{1}(h)}$ is mapped
by (\ref{glue1}) into a wedge of angular opening $\frac{2\pi }{3}$. This is
consistent with the definition of $|P_{T_{L}}|\rightarrow {M}$ according to
which the vertices $\{\rho ^{0}(j)\}\in |P_{T_{L}}|\rightarrow {M}$ are the
barycenters of $\ $the flat $\{\sigma ^{2}(j)\}\in |T_{l}|\rightarrow M$. \
\ Similarly, if $\{U_{\rho ^{1}(h_{\beta })}\}$, $\beta =1,2,...,q(k)$ are
the open strips associated with the $q(k)$ (oriented) edges $\{\rho
^{1}(h_{\beta })\}$ boundary of the generic polygonal cell $\rho ^{2}(k)$,
then the transition functions between the corresponding uniformizing
coordinate $\zeta (k)$ and the $\{z(h_{\beta })\}$ are given by [20] 
\begin{equation}
\zeta (k)=\exp \left( \frac{2\pi i}{L(\partial (\rho ^{2}(k)))}\left(
\sum_{\beta =1}^{\nu -1}L(\rho ^{1}(h_{\beta }))+z(h_{\nu })\right) \right)_{\nu
=1}^{q(k)},  \label{glue2}
\end{equation}
with $\sum_{\beta =1}^{\nu -1}\cdot \doteq 0$, for $\nu =1$.

Summing up, we have the following result which can be considered as a rather
elementary consequence of the connection between ribbon graphs and the
complex analytic theory of Teichm\"{u}ller spaces (see [20]),

\begin{proposition}
The mapping 
\begin{gather}
\Upsilon :(|P_{T_{L}}|\rightarrow {M)\longrightarrow }((M;N_{0}),\mathcal{C})
\label{riemsurf} \\
\Gamma \longmapsto \bigcup_{\{\rho ^{0}(j)\}}^{N_{2}(T)}U_{\rho
^{0}(j)}\bigcup_{\{\rho ^{1}(h)\}}^{N_{1}(T)}U_{\rho ^{1}(h)}\bigcup_{\{\rho
^{2}(k)\}}^{N_{0}(T)}U_{\rho ^{2}(k)},  \notag
\end{gather}
\ \ where the glueing maps are given by (\ref{glue1}) and (\ref{glue2}), \
defines the (punctured) Riemann surface $((M;N_{0}),\mathcal{C})$ \
canonically associated with the conical Regge polytope $|P_{T_{L}}|%
\rightarrow {M}$.
\end{proposition}

\noindent Note that by construction such a Riemann surface carries the decoration
provided by the meromorphic quadratic differential $\psi $. It is through
such a decoration that the punctured Riemann surface $((M;N_{0}),\mathcal{C}%
,\psi )$ keeps track of the metric geometry of the conical Regge polytope $%
|P_{T_{L}}|\to {M}$ out of which $((M;N_{0}),\mathcal{C})$\ has been
generated. Explicitly, since the meromorphic quadratic differential (\ref
{differ}) has a second order poles, the correspondence $\Upsilon $ defined
by (\ref{riemsurf}) associates with the generic two-cell $\rho ^{2}(k)\in $ $%
|P_{T_{L}}|\rightarrow {M}$, a punctured disk 
\begin{equation}
\Delta _{k}^{\ast }\doteq \{\zeta (k)\in \mathbb{C}|\;0<|\zeta (k)|<1\}
\end{equation}
endowed with a flat metric 
\begin{equation}
|\psi (k)_{\rho ^{2}(k)}|=\frac{\left[ L(\partial (\rho ^{2}(k)))\right] ^{2}%
}{4\pi ^{2}|\zeta (k)|^{2}}|d\zeta (k)|^{2}.  \label{flmetr}
\end{equation}
On $(\Delta _{k}^{\ast },|\psi (k)_{\rho ^{2}(k)}|)$ \ \ we can evaluate the
length $L(\Upsilon (l))$ of any closed curve $l$, homotopic to $\partial
(\rho ^{2}(k))$ and contained in $\rho ^{2}(k)-\sigma ^{0}(k)$, 
\begin{equation}
L(\Upsilon (l))\doteq \oint_{l\sim \partial (\rho ^{2}(k))}\sqrt{\psi
(k)_{\rho ^{2}(k)}}=L(\partial (\rho ^{2}(k))).  \label{length}
\end{equation}
\ \ Moreover, if we define $\Delta _{k}^{\varrho }\doteq \left\{ \zeta
(k)\in \mathbb{C}|\;\varrho <|\zeta (k)|<1\right\} $, then in terms of the \
area element $|\psi (k)|d\func{Re}(\zeta (k))\wedge $\ $d\func{Im}(\zeta
(k)) $ associated with the flat metric $|\psi (k)_{\rho ^{2}(k)}|$ we get 
\begin{equation}
\int_{\Delta _{k}^{\varrho }}|\psi (k)_{\rho ^{2}(k)}|d\func{Re}(\zeta
(k))\wedge \ d\func{Im}(\zeta (k))=\frac{\left[ L(\partial (\rho ^{2}(k)))%
\right] ^{2}}{2\pi }\ln \left( \frac{1}{\varrho }\right) ,  \label{area}
\end{equation}
which, as $\varrho \rightarrow 0^{+}$, diverges logarithmically.\ \ Thus,
from a metrical point of view the punctured disk $(\Delta _{k}^{\ast },|\psi
(k)_{\rho ^{2}(k)}|)$,\emph{\ }endowed with the flat metric $|\psi (k)_{\rho
^{2}(k)}|$, is isometric to a flat semi-infinite cylinder. In this
connection, it is interesting to remark that the flat metric (\ref{flmetr})
formally corresponds to the limiting case of the conical metric (\ref{cmetr}%
) when the total angle of the cone $\theta (k)\rightarrow 0$, thus
interpreting the flat semi-infinite cylinder as a degenerate cone.

\bigskip

\noindent The explicit connection between the singular uniformization $\left( M,%
\mathcal{C}_{sg}\right) $ associated with the singular Euclidean structure (%
\ref{cmetr}), (\ref{sing}) and the decoration $((M;N_{0}),\mathcal{C},\psi )$
of $((M;N_{0}),\mathcal{C})$ associated with the quadratic differential $%
\psi $ can be easily worked out and is defined (up to a normalization) by
the mapping 
\begin{gather}
(t_{k},D_{k}(r(k)))\longrightarrow (\zeta (k),U_{\rho ^{2}(k)})
\label{tzeta} \\
t_{k}\longmapsto \zeta (k)=\exp \frac{2\pi }{L(\partial (\rho ^{2}(k)))}%
\left[ \frac{2\pi }{2\pi -\varepsilon (k)}\left( t_{k}-t_{k}(\sigma
^{0}(k))\right) ^{\frac{2\pi -\varepsilon (k)}{2\pi }}\right]  \notag
\end{gather}
for each $k=1,...,N_{0}(T)$.

\bigskip

The picture of the correspondence (\ref{riemsurf})\ \ which emerges from
such an analysis the following: the map $\Upsilon $ associates with a
conical Regge polytope $|P_{T_{L}}|\rightarrow {M}$ the pair $(\mathcal{C}%
,\psi )$, where $\mathcal{C}$ is a complex structure on $(M;N_{0})$. Note
that since $2-2g-N_{0}(T)<0$, the punctured Riemann surface $(M;N_{0},%
\mathcal{C})$\ can be endowed with a hyperbolic structure, \emph{i.e.}, with
a metric $ds_{-1}^{2}[\mathcal{C}]$ in the conformal class of $(M;N_{0},%
\mathcal{C})$ which is complete on $(M;N_{0},\mathcal{C})$ and has constant
curvature $-1$. Explicitly, on each punctured disk $\Delta _{k}^{\ast }$
associated with two-cell $\rho ^{2}(k)\in $ $|P_{T_{L}}|\rightarrow {M}$, we
can locally write 
\begin{equation}
\left. ds_{-1}^{2}[\mathcal{C}]\right| _{\Delta _{k}^{\ast }}=\left[ \frac{%
\left| d\zeta (k)\right| }{|\zeta (k)|\ln |\zeta (k)|}\right] ^{2},
\label{hyper}
\end{equation}
an expression which can be easily obtained by quotienting he upper
half-plane equipped with the standard hyperbolic metric $(\ \mathcal{H}%
^{2},h(\zeta )^{2}\left| d\zeta \right| ^{2})$ by the isometry $\zeta
\rightarrow \zeta +1$ and by identifying the resulting quotient with (the
unit punctured disk) $\Delta _{k}^{\ast }$ via the map $\zeta \rightarrow
\exp [2\pi \sqrt{-1}\zeta ]$. According to these remarks, the natural
complex structure $(M;N_{0},\mathcal{C})$ dual to (\ref{cmetr}) and
associated with the conical Regge polytope $|P_{L}|\rightarrow M$ can be
obtained by glueing to the boundary components of the ribbon graph $\Gamma $
a corresponding set of punctured disks $(\Delta _{k}^{\ast },\left.
ds_{-1}^{2}[\mathcal{C}]\right| _{\Delta _{k}^{\ast }})$ endowed with the
hyperbolic metric (\ref{hyper}) and decorated with the quadratic
differential $\psi (k)_{\rho ^{2}(k)}$. The punctures are identified with
the vertices $\{\sigma ^{0}(j)\}$\ of the Regge triangulation $%
|T_{l}|\rightarrow M$ which, upon barycentrical dualization, gives rise to $%
|P_{T_{L}}|\rightarrow {M}$. Thus, pictorially, the mapping \ $\Upsilon $
establishes the following correspondence 
\begin{equation}
\left\{ 
\begin{tabular}{c}
Space of singular \\ 
Euclidean Structures \\ 
on $M$ with $N_{0}(T)$ \\ 
conical points
\end{tabular}
\right\} \Longrightarrow \left\{ 
\begin{tabular}{c}
Decorated \\ 
Punctured \\ 
Riemann Surfaces \\ 
on $(M,N_{0})$%
\end{tabular}
\right\} 
\end{equation}

Not unexpectedly, the set of (generalized) conical Regge polytopes
(trivalent ribbon graphs) can be used to combinatorially parametrize, by
means of $(M;N_{0},\mathcal{C})$, the moduli space $\mathfrak{M}_{g},_{N_{0}}$
of genus $g$ Riemann surfaces\ with $N_{0}$ punctures, (see the Appendix).
We are being a bit vague here, but\ we will comment on some aspect of this
well-known parametrization shortly.

\bigskip

From a topological viewpoint, the set of all possible metrics $|\psi |$ on a
(trivalent) ribbon graph $\Gamma $ with given edge-set $e(\Gamma )$\ can be
characterized [20], [21] as a space homeomorphic to $\mathbb{R}%
_{+}^{|e(\Gamma )|}$, ($|e(\Gamma )|$ denoting the number of edges in $%
e(\Gamma )$), topologized by the \ standard $\epsilon $-neighborhoods \ $%
U_{\epsilon }\subset $ $\mathbb{R}_{+}^{|e(\Gamma )|}$. On such a space
there is a natural action of $Aut(\Gamma )$, the automorphism group of $%
\Gamma $ defined by the homomorphism $Aut(\Gamma )\rightarrow \mathfrak{G}%
_{e(\Gamma )}$ where $\mathfrak{G}_{e(\Gamma )}$ denotes the symmetric group
over $|e(\Gamma )|$ elements. Thus, the resulting space $\mathbb{R}%
_{+}^{|e(\Gamma )|}/Aut(\Gamma )$ is a differentiable orbifold. Let 
\begin{equation}
Aut_{\partial }(P_{L})\subset Aut(P_{L}),
\end{equation}
denote the subgroup of ribbon graph automorphisms of \ the (trivalent) $1$%
-skeleton $\Gamma $ of \ $|P_{T_{L}}|\rightarrow {M}$\ that preserve the
(labeling of the) boundary components of $\Gamma $. Then, the space of \ $1$%
-skeletons of conical\ Regge polytopes\ $|P_{T_{L}}|\rightarrow {M}$, with $%
N_{0}(T)$ labelled boundary components, on a surface $M$ of genus $g$ can be
defined by [20] 
\begin{equation}
RGP_{g,N_{0}}^{met}=\bigsqcup_{\Gamma \in RGB_{g,N_{0}}}\frac{\mathbb{R}%
_{+}^{|e(\Gamma )|}}{Aut_{\partial }(P_{L})},  \label{DTorb}
\end{equation}
where the disjoint union is over the subset of all trivalent ribbon graphs
(with labelled boundaries) satisfying the topological stability condition $%
2-2g-N_{0}(T)<0$, and which are dual to generalized triangulations. It
follows, (see [20] theorems 3.3, 3.4, and 3.5), that the set $%
RGP_{g,N_{0}}^{met}$ is locally modelled on a stratified space constructed
from the components (rational orbicells) $\mathbb{R}_{+}^{|e(\Gamma
)|}/Aut_{\partial }(P_{L})$ by means of a (Whitehead) expansion and collapse
procedure for ribbon graphs, which basically amounts to collapsing edges and
coalescing vertices, (the Whitehead move in $|P_{T_{L}}|\rightarrow {M}$ is
the dual of the familiar flip move for triangulations). Explicitly, if $%
L(t)=tL$ is the length of an edge $\rho ^{1}(j)$ of a ribbon graph $\Gamma
_{L(t)}\in $ $RGP_{g,N_{0}}^{met}$, then, as $t\rightarrow 0$, we get the
metric ribbon graph $\widehat{\Gamma }$ which is obtained from $\Gamma
_{L(t)}$ by collapsing the edge $\rho ^{1}(j)$. By exploiting such
construction, we can extend the space $RGP_{g,N_{0}}^{met}$ to a suitable
closure $\overline{RGP}_{g,N_{0}}^{met}$ [21].\ \ Since the dual of any
metric ribbon graph $\in RGP_{g,N_{0}}^{met}$ \ is a (generalized)
triangulation $|T_{_{L}}|\rightarrow {M}$ of the surface $M$, we have the
following equivalent characterization [21]

\begin{remark}
The rational orbicells $\frac{{R}_{+}^{|e(\Gamma )|}}{Aut_{\partial }(P_{L})}
$ of \ $RGP_{g,N_{0}}^{met}$\ can be labelled by the generalized
triangulations associated with the generalized conical Regge polytopal
surfaces $|P_{T_{L}}|\rightarrow {M}$.
\end{remark}

The open cells of \ $RGP_{g,N_{0}}^{met}$, being associated with trivalent
graphs, have dimension provided by the number $N_{1}(T)$ of edges of $%
|P_{T_{L}}|\rightarrow {M}$. From the Euler relation $%
N_{0}(T)-N_{1}(T)+N_{2}(T)=2-2g$, and the constraint $2N_{1}(T)=3N_{2}(T)$
associated with the trivalency, we get 
\begin{equation}
\dim \left[ RGP_{g,N_{0}}^{met}\right] =N_{1}(T)=3N_{0}(T)+6g-6.
\end{equation}
There is a natural projection 
\begin{gather}
p:RGP_{g,N_{0}}^{met}\longrightarrow \mathbb{R}_{+}^{N_{0}(T)} \\
\Gamma \longmapsto p(\Gamma )=(L_{1},...,L_{N_{0}(T)}),  \notag
\end{gather}
where $(L_{1},...,L_{N_{0}(T)})$ denote the perimeters of the polygonal
2-cells $\{\rho ^{2}(j)\}$ of $|P_{T_{L}}|\rightarrow {M}$. With respect to
the topology on the space of metric ribbon graphs, the orbifold $%
RGP_{g,N_{0}}^{met}$ endowed with such a projection acquires\ the structure
of a cellular bundle. For a given sequence $\{L(\partial (\rho ^{2}(k)))\}$,
the fiber 
$$p^{-1}(\{L(\partial (\rho ^{2}(k)))\})=$$
\begin{equation}
=\left\{ |P_{T_{L}}|\rightarrow {M}\in
RGP_{g,N_{0}}^{met}:\{L_{k}\}=\{L(\partial (\rho ^{2}(k)))\}\right\}
\end{equation}
is the set of all generalized conical Regge polytopes with the given 
perimeters. In particular, the equilateral conical Regge polytopes dual to
the set \ of \ distinct generalized dynamically triangulated surfaces with
given curvature assignments $\{q(i),\sigma ^{0}(i)\}_{i=1}^{N_{0}(T)}$, is
contained in
\begin{equation}
p^{-1}\left\{ \{L(\partial (\rho ^{2}(k)))\}=\frac{\sqrt{3}}{3}%
a\{q(k)\}_{k=1}^{N_{0}}\right\} .
\end{equation}
If we take into account the $N_{0}(T)$ constraints associated with the
perimeters assignments, it follows that the fibers $p^{-1}(\{L(\partial
(\rho ^{2}(k)))\})$ have dimension provided by 
\begin{equation}
\dim \left[ p^{-1}(\{L(\partial (\rho ^{2}(k)))\}\right] =2N_{0}(T)+6g-6,
\end{equation}
which, as is well known, exactly corresponds to the real dimension of the
moduli space $\mathfrak{M}_{g},_{N_{0}}$ of genus $g$ Riemann surfaces\ with $%
N_{0}$ punctures.\ \ \ 

We conclude this section by observing that corresponding to each marked
polygonal 2-cells $\{\rho ^{2}(k)\}$ of $|P_{T_{L}}|\rightarrow {M}$ there
is a further (combinatorial) bundle map 
\begin{equation}
\mathcal{CL}_{k}\rightarrow RGP_{g,N_{0}}^{met}
\end{equation}
whose fiber over $(\Gamma ,\rho ^{2}(1),...,\rho ^{2}(N_{0}))$ is provided
by the boundary cycle $\partial \rho ^{2}(k)$, (recall that each boundary $%
\partial \rho ^{2}(k)$ comes with a positive orientation). To any such $%
\mathcal{CL}_{k}$ one associates [8], [21] the piecewise smooth 2-form
defined by 
\begin{equation}
\omega _{k}(\Gamma )=\sum_{1\leq h_{\alpha }<h_{\beta }\leq q(k)-1}d\left( 
\frac{L(\rho ^{1}(h_{\alpha }))}{L(\partial \rho ^{2}(k))}\right) \wedge
d\left( \frac{L(\rho ^{1}(h_{\beta }))}{L(\partial \rho ^{2}(k))}\right) ,
\label{chern}
\end{equation}
which is invariant under rescaling and cyclic permutations of the $L(\rho
^{1}(h_{\mu }))$ and represents the first Chern class of $\mathcal{CL}_{k}$.

\section{DT and moduli spaces: the nature of the approximation in simplicial
quantum gravity}

\bigskip

\ \ \ As already stressed, the above analysis can be considered as a
particular case of the well-known results which allows to define a bijective
mapping (a homeomorphism of orbifolds) between the space of ribbon graphs $%
RGB_{g,N_{0}}^{met}$ and the moduli space $\mathfrak{M}_{g},_{N_{0}}$ of genus $g
$ Riemann surfaces $((M;N_{0}),\mathcal{C})$ with $N_{0}(T)$\ punctures
[20], [21], 
\begin{gather}
h:\mathfrak{M}_{g},_{N}\times {R}_{+}^{N}\rightarrow RGB_{g,N_{0}}^{met}
\label{bijec} \\
\lbrack ((M;N_{0}),\mathcal{C}),L_{i}]\longmapsto \Gamma ,  \notag
\end{gather}
where $(L_{1},...,L_{N_{0}})$ is an ordered n-tuple of positive real numbers
and $\Gamma $ is a metric ribbon graphs with $N_{0}(T)$ labelled boundary
lengths $\{L_{i}\}$ defined by the corresponding JS quadratic differential.
The bijection $h$ extends to $\overline{\mathfrak{M}}_{g},_{N_{0}}\times {R}%
_{+}^{N_{0}}\rightarrow \overline{RGB}_{g,N_{0}}^{met}$ in such a way that
two (stable) surfaces $M_{1}$ and $M_{2}$ with $N_{0}(T)$ punctures and
given perimeters $\{L_{i}\}$ are mapped in the same ribbon graph $\Gamma \in 
\overline{RGB}_{g,N_{0}}^{met}$ if and only if there exists an homeomorphism
between $M_{1}$ and $M_{2}$ preserving the (labelling of the) punctures, and
is holomorphic on each irreducible component containing one of the
punctures. If one looks at moduli spaces from this point of view, the
bundles $\mathcal{CL}_{k}$ over $\overline{RGP}_{g,N_{0}}^{met}$ defined in
the previous section are the natural combinatorial counterpart of the line
bundles $\mathcal{L}_{k}\rightarrow \overline{\mathfrak{M}}_{g},_{N_{0}}$ whose
fiber at the moduli point $((M;N_{0}),\mathcal{C})$ is defined by the
cotangent space $T_{(M,p_{k})}^{\ast }$ to $((M;N_{0}),\mathcal{C})$ at $%
p_{k}\doteq \sigma ^{0}(k)$. In particular, the pull-back under (\ref{bijec}%
) of the $2$-form $\omega _{k}(\Gamma )$ defined by (\ref{chern}) is a
(combinatorial) representative of the first Chern class $c_{1}(\mathcal{L}%
_{k})$ of the line bundle $\mathcal{L}_{k}\rightarrow \overline{\mathfrak{M}}%
_{g},_{N_{0}}$.

\bigskip

\noindent In full generality, the construction in [20] (see \S \S 4 and 5) to which we
refer for details,\ gives rise to an explicit map from the whole space of
metric ribbon graphs $RGB_{g,N_{0}}^{met}$ to the decorated moduli space $%
\mathfrak{M}_{g},_{N}\times {R}_{+}^{N}$. The explicit expression for such a
mapping in the case of the trivalent graph associated with a conical Regge
polytope is provided by (\ref{riemsurf}) and can be easily specialized to
the DT framework according to

\begin{proposition}
Let 
$$\mathcal{DT}[\{q(k)\}_{k=1}^{N_{0}}]\doteq$$
\begin{equation}
\doteq\left\{ |T_{l=a}|\rightarrow
M\;:q(\sigma ^{0}(k))=q(k)\geq 2,\;\;k=1,...,N_{0}(T)\right\} .
\end{equation}
\ \ denote the set of \ distinct generalized dynamically triangulated
surfaces of genus $g$, with a given set of ordered curvature assignments $%
\{q(i),\sigma ^{0}(i)\}_{i=1}^{N_{0}(T)}$ over its $N_{0}(T)$ labelled
vertices. Assume that the topological stability condition $g\geq 0$, $%
2-2g-N_{0}(T)<0$ holds. Then, we can associate with each polygonal $2$-cell $%
\{\rho ^{2}(i)\}_{i=1}^{N_{0}(T)}$ of the conical polytope $%
|P_{T_{a}}|\rightarrow {M}$ dual to a $|T_{l=a}|\rightarrow M \in \mathcal{DT}%
[\{q(k)\}_{k=1}^{N_{0}}]$ the quadratic differential 
\begin{equation}
\rho ^{2}(k)\longmapsto \psi (k)\doteq -\frac{\left( \frac{\sqrt{3}}{3}%
aq(k)\right) ^{2}}{4\pi ^{2}\zeta ^{2}(k)}d\zeta (k)\otimes d\zeta (k),
\label{quadiff}
\end{equation}
\ \ \ where $\zeta (k)$ is a locally uniformizing complex coordinate in the
unit disk. The set of such $\{\psi (k),\zeta (k)\}_{k=1}^{N_{0}(T)}$
naturally restricts to the edges $\{\rho ^{1}(j)\}_{j=1}^{N_{1}(T)}$ of \ $%
|P_{T_{a}}|\rightarrow {M}$ and characterizes uniquely a complex structure $%
\mathcal{C}$ \ and a meromorphic quadratic differential $\psi d\zeta \otimes
d\zeta $\ which define a decorated punctured Riemann surface $(((M;N_{0}),%
\mathcal{C}),\psi )$ associated with $|T_{l=a}|\rightarrow M$. It follows
that there is an injective mapping \ \ 
\begin{gather}
f_{T_{a}}:\mathcal{DT}[\{q(k)\}_{k=1}^{N_{0}}]\longrightarrow \mathfrak{M}%
_{g},_{N_{0}}\times \left( \frac{\sqrt{3}}{3}a\right) {N}_{+}^{N_{0}}
\label{injection} \\
\left( |T_{l=a}|\rightarrow M;\{q(k)\}_{k=1}^{N_{0}}\right) \longmapsto
\lbrack ((M;N_{0}),\mathcal{C}),\psi ],  \notag
\end{gather}
defining the generalized dynamical triangulations $|T_{l=a}|\rightarrow M\in\mathcal{DT}[\{q(k)\}]$ as distinguished labellings (i.e., providing a
reference complex structure $\mathcal{C}_{T_{a}})$ in the top-dimensional
orbicells 
$$h^{-1}\left( \frac{{R}_{+}^{|e(\Gamma )|}}{Aut_{\partial }(P_{L})}%
\right) \subset \mathfrak{M}_{g},_{N_{0}}\times {R}_{+}^{N}.$$
\end{proposition}

\emph{Proof}. In the light of the above remarks, very little remains at
issue here. The first part of the proposition is a trivial consequence of
the explicit construction of \ the mapping (\ref{differ}) $%
h^{-1}:RGB_{g,N_{0}}^{met}\rightarrow \mathfrak{M}_{g},_{N}\times {R}_{+}^{N}$ \
for the trivalent ribbon graphs associated with \ $|P_{T_{L}}|\rightarrow {M}
$. \ In order to prove the second part let us denote by 
\begin{equation}
\Omega _{T_{a}}(\{q(k)\}_{k=1}^{N_{0}})\doteq \frac{{R}_{+}^{|e(\Gamma )|}}{%
Aut_{\partial }(P_{T_{a}})}  \label{omega}
\end{equation}
the rational cell associated with the dynamical triangulation $%
|T_{l=a}|\rightarrow M\in \mathcal{DT}[\{q(k)\}_{k=1}^{N_{0}}]$. Note that
the automorphism group $Aut_{\partial }(P_{T_{a}})$ is isomorphic to the
isotropy subgroup $\subset \mathfrak{Map}(M;N_{0})$ of the generic Riemann
surface in $h^{-1}(\Omega _{T_{a}}(\{q(k)\}_{k=1}^{N_{0}}))$. The orbicell (%
\ref{omega}) contains the ribbon graph associated with the conical polytope $%
|P_{T_{a}}|\rightarrow {M}$ dual to $|T_{l=a}|\rightarrow M$, and all
(trivalent) metric ribbon graphs $|P_{T_{L}}|\rightarrow {M}$\ with the same
combinatorial structure of \ $|P_{T_{a}}|\rightarrow {M}$ but with all
possible length assignments $\{L(\rho ^{1}(h))\}_{1}^{N_{1}(T)}$ associated
with the corresponding set of edges $\{\rho ^{1}(h)\}_{1}^{N_{1}(T)}$. Up to
the action of \ $Aut_{\partial }(P_{T_{a}})$,\ \ (\ref{omega}) is naturally
identified \ with the convex polytope (of dimension $(2N_{0}(T)+6g-6)$) in $%
\mathbb{R}_{+}^{N_{1}(T)}$ defined by 
\begin{equation}
\Delta _{T_{a}}(\{q(k)\}_{k=1}^{N_{0}})=  \notag
\end{equation}
\begin{equation}
=\left\{ \{L(\rho ^{1}(j))\}\in \mathbb{R}_{+}^{N_{1}(T)}:\left.
\sum_{j=1}^{q(k)}A_{(k)}^{j}(T_{a})L(\rho ^{1}(j))=\frac{\sqrt{3}}{3}%
aq(k)\right| _{k=1}^{N_{0}(T)}\;\right\},\notag  
\end{equation}
where $A_{(k)}^{j}(T_{a})$ is a $(0,1)$ indicator matrix, depending on the
given dynamical triangulation $|T_{l=a}|\rightarrow M$, with $%
A_{(k)}^{j}(T_{a})=1$ if the edge $\rho ^{1}(j)$ belongs to $\partial (\rho
^{2}(k))$, and $0$ otherwise. Note that $|P_{T_{a}}|\rightarrow {M}$ appears
as the barycenter of such a polytope.

Since the cell decomposition (\ref{DTorb}) of the space of trivalent metric
ribbon graphs $RGP_{g,N_{0}}^{met}$ depends only on the combinatorial type
of the ribbon graph, we can use the equilateral polytopes $%
|P_{T_{a}}|\rightarrow {M}$, dual to dynamical triangulations in $\mathcal{DT%
}[\{q(k)\}_{k=1}^{N_{0}}]$ , as the set over which the disjoint union in (%
\ref{DTorb}) runs. Thus we can write 
\begin{equation}
RGP_{g,N_{0}}^{met}=\bigsqcup_{\mathcal{DT}[\{q(k)\}_{k=1}^{N_{0}}]}\Omega
_{T_{a}}(\{q(k)\}_{k=1}^{N_{0}}).
\end{equation}
\ \ \ Note that the decoration associated with the quadratic differential \ $%
\psi $ is provided by

\begin{equation}
\oint_{\partial (\rho ^{2}(k))}\sqrt{\psi (k)}=\sum_{h=1}^{q(k)}L(\rho
^{1}(h))\underset{DT}{=}\left( \frac{\sqrt{3}}{3}a\right) q(k).
\label{perim}
\end{equation}

\bigskip

\begin{remark}
In the case of dynamical triangulations the above integral can be actually
exploited for providing the conical angles $\theta (k)=$ $\frac{\pi }{3}q(k)$
associated with the dynamical triangulation $|T_{l=a}|\rightarrow M$. In
such a sense, the decoration defined by the quadratic differentials $\{\psi
(k)\}$ allows for a reconstruction of the original triangulation $%
|T_{l=a}|\rightarrow M$ characterizing the orbicell $\Omega
_{T_{a}}(\{q(k)\}_{k=1}^{N_{0}})$. Thus, the quadratic differential $\psi
(k) $ can be interpreted as a field on the Riemann surface $((M;N_{0}),%
\mathcal{C})$ associated with $|T_{l=a}|\rightarrow M$ and carrying
curvature. It couples with the punctures by dressing them with a
gravitational charge provided by (\ref{perim}).
\end{remark}

\bigskip

Roughly speaking the above remarks imply that we can view dynamical
triangulations as punctured Riemann surfaces dressed by a discretized
curvature operator. The distinguished role that dynamical triangulations
play in the correspondence (\ref{bijec}) \ may appear a little bit
mysterious and its worthwhile discussing its origin. The starting point is
the observation that there is a particular simplicial refinement of a given
Regge triangulation $|T_{l}|\rightarrow M$ which, at the expenses of adding
new regular (\emph{i.e.}, with $\theta (k)=2\pi $) vertices generates from $%
|T_{l}|\rightarrow M$ an almost-equilateral triangulations $|T_{l\approx
a}|\rightarrow M$ characterized by the same (essential) divisor $Div(T)$ of
the original triangulation $|T_{l}|\rightarrow M$. The refinement in
question is the so called \emph{hex refinement} familiar in the theory of
circle packing, [22], [23], [24]. The hex refinement $HX\{|T_{l}|\rightarrow
M\}$ of $|T_{l}|\rightarrow M$ is the triangulation formed by adding a
vertex to each edge of $|T_{l}|\rightarrow M$ and adding an edge between new
vertices lying on the same triangular face. In this way each face of \ $%
|T_{l}|\rightarrow M$ will be subdivided into four new faces, and any new
interior vertex added to $|T_{l}|\rightarrow M$ is a flat regular vertex (of
degree $q(k)=6$ and $\theta (k)=2\pi $). The original conical angles $%
\{\theta (k)\}$ of $|T_{l}|\rightarrow M$ can be, at each step $\left(
HX\{|T_{l}|\rightarrow M\}\right) ^{n}$, changed by retaining the original
incidence numbers $\{q(k)\}$, and eventually ($n\rightarrow \infty $)
fine-tuned to $\theta (k)\rightsquigarrow q(k)\frac{\pi }{3}$ .\ As a matter
of fact, by iterating such a procedure it is possible to prove [25] that the
resulting triangulation $\left( HX\{|T_{l}|\rightarrow M\}\right) ^{n}$
converges to the triangulated surface formed by glueing together equilateral
triangles in the pattern given by the original $|T_{l}|\rightarrow M$.\ The
Riemann surface associated with $\left( HX\{|T_{l}|\rightarrow M\}\right)
^{n}$ correspondingly converges, in the Teichm\"{u}ller metric (see the
appendix, (\ref{teichnorm})), to the surface associated with the equilateral
triangulation.

This property of dynamically triangulated surfaces is strictly related to a
result, due to V. A. Voevodskii and G. B. Shabat [26], which establishes a
remarkable bijection between dynamical triangulations and curves over
algebraic number fields. The proof in [26] exploits the characterization of
the collection of algebraic curves defined over the\ algebraic closure $%
\overline{\mathbb{Q}}$ of the field of rational numbers,\ (\emph{i.e.}, over
the set of complex numbers which are roots of non-zero polynomials with
rational coefficients), provided by Bely\u{\i}'s theorem (see e.g., [20]).
According to such a result, a nonsingular Riemann surface $M$ has the
structure of an algebraic curve defined over $\overline{\mathbb{Q}}$ if and
only if there is a holomorphic map (a branched covering of $M$ over the
sphere) 
\begin{equation}
f:M\rightarrow \mathbb{CP}^{1}
\end{equation}
that is ramified only at $0,1$ and $\infty $, (such maps are known as Bely\u{%
\i} maps). In other words, the Riemann surface associated with a dynamical
triangulation (with edge-lengths normalized to $a=1$) is, in a canonical
way, \ a ramified covering of the Riemann sphere $\mathbb{CP}^{1}$ with
ramification locus contained in $\{0,1,\infty \}$. The triangulation
(actually its barycentrically dual ribbon graph decorated with the quadratic
differential $\psi $) is the preimage of the interval $[0,1]$, in particular
the set of vertices appears as the preimage of $0$ and the set of
half-cylinders over the cells $\{\rho ^{2}(k)\}$ as the preimage of $\infty $%
. The mid points of the edges $\{\rho ^{1}(h)\}$, (identified with the
barycenters of the edges of \ $|T_{a=1}|\rightarrow M$), correspond to the
preimage of $1$. Moreover, every branched covering of $\mathbb{CP}^{1}$
defining a Riemann surface over $\overline{\mathbb{Q}}$ arises in this
fashion.\ It is worthwhile remarking that the inverse image of the line
segment $[0,1]\subset \mathbb{CP}^{1}$ under a Bely\u{\i} map is a
Grothendieck's \emph{dessin d'enfant}, thus dynamically triangulated
surfaces are eventually connected with the theory of the Galois group $Gal(%
\overline{\mathbb{Q}}/\mathbb{Q})$ action on the branched coverings $%
f:M\rightarrow \mathbb{CP}^{1}$. The correspondence between Bely\u{\i} maps, 
\emph{dessin d'enfant} and JS quadratic differentials has been recently
analyzed in depth by M. Mulase and M. Penkava [20], an equally inspiring
paper is [27] by M. Bauer and C. Itzykson.

\begin{remark}
Since Bely\u{\i} surfaces are dense in the moduli space of Riemann surfaces,
such density carries over to the set of dynamically triangulated surfaces.
\end{remark}

\noindent In this connection, we are actually interested in understanding to what
extent the map $f_{T_{a}}$ (\ref{injection}) and more generally $%
RGP_{g,N_{0}}^{met}\rightarrow \mathfrak{M}_{g},_{N_{0}}\times {R}_{+}^{N_{0}}$
is an approximation to the conformal parametrization of the space of
Riemannian structures on a surface $M$, used in the standard field-theoretic
approach to $2D$ quantum gravity. To begin with, let us remark that the
above analysis provides a formal dictionary between the smooth and the
combinatorial description of a Riemannian surface $(M,ds^{2})$ according to 
\begin{equation}
\begin{tabular}{c}
Riemannian Surfaces \\ 
\\ 
$\frac{Riem(M)}{Diff(M)}$ \\ 
\\ 
$\mathfrak{M}_{g}$ \\ 
\\ 
$W(M)$%
\end{tabular}
\begin{tabular}{c}
\\ 
\\ 
$\rightsquigarrow \rightsquigarrow $ \\ 
\\ 
$\hookrightarrow \longrightarrow $ \\ 
$\nearrow $ \\ 
$\longrightarrow \longrightarrow $%
\end{tabular}
\begin{tabular}{c}
Triangulations \\ 
\\ 
$RGP_{g,N_{0}}^{met}$ \\ 
\\ 
$\mathfrak{M}_{g},_{N_{0}}$ \\ 
\\ 
$\mathbb{R}_{+}^{N_{0}}$%
\end{tabular}
\label{diction}
\end{equation}
Where$\ \frac{Riem(M)}{Diff(M)}\rightsquigarrow $\ $RGP_{g,N_{0}}^{met}$\
stands for the approximation of a two dimensional Riemannian structure defined
by a dynamical triangulation and 
\begin{gather}
\mathfrak{M}_{g}\hookrightarrow \mathfrak{M}_{g},_{N_{0}} \\
W(M)\ \overset{\nearrow }{\longrightarrow }\mathbb{R}_{+}^{N_{0}}  \notag
\end{gather}
represents a correspondence between $\mathfrak{M}_{g},_{N_{0}}$, the moduli
space of Riemann surfaces of genus $g$, $\mathfrak{M}_{g}$, and the
discretization of the conformal degrees of freedom $W(M)$. The nature of
such a correspondence is rather non-trivial and its analysis will occupy us
in the remaining part of this paper. First of all, it cannot be explained if
we simply regard dynamical triangulations as providing a sort of \
approximating net in the space of Riemannian structures $\frac{Riem(M)}{%
Diff(M)}$.\ Actually, the rationale we are seeking lies at a deeper level
and is a direct consequence of the interpretation of dynamical triangulation
as punctured Riemann surfaces dressed by a discretized curvature, (see
remark 8 ). \ If we approximate a Riemannian structure with a (dynamical)
triangulation, $(M,ds^{2})\rightsquigarrow (|T_{l}|\rightarrow M)$, then the
metric information $ds_{T}^{2}$ of the approximating $|T_{l}|\rightarrow M$
is encoded into the lengths of the $N_{1}(T)$ edges of the triangulation.
This translates into $6g-6+3N_{0}(T)$ parameters, as we have seen in section
2, where we wrote $ds_{T}^{2}=e^{2v}|dt|^{2}$ with the conformal factor
locally provided by $v\doteq u-\sum_{k=1}^{N_{0}(T)}\left( \frac{\varepsilon
(k)}{2\pi }\right) \ln \left| t-t_{k}\right| $. Thus, the conformal mode is
approximated by $3N_{0}(T)$ parameters. However, out of these $3N_{0}(T)$
quantities, the $N_{0}(T)$ deficit angles $\{\varepsilon (k)\}$ (\emph{i.e.}
the curvatures) are basically free variables being subjected only to the
linear topological constraint \ $\sum_{k=1}^{N_{0}(T)}\left( -\frac{%
\varepsilon (k)}{2\pi }\right) =2g-2$, (this follows directly from the
Picard-Troyanov theorem). More delicate is the situation of the remaining $%
2N_{0}(T)$ degrees of freedom associated with the (complex) coordinates $%
t_{k}$ of the conical singularities (vertices), and described by $\ln \left|
t-t_{k}\right| $. These latter conformal degrees of freedom basically
describe the conical points where the curvature of $|T_{l}|\rightarrow M$ \
couples with the underlying complex structure of \ $(M,ds_{T}^{2})$ and
cannot be easily handled (see  [7]). However, if we go to the
barycentrically dual complex associated with $(|T_{l}|\rightarrow M)$, then
the $2N_{0}(T)$ conformal degrees of freedom associated with the (complex)
coordinates $t_{k}$ get naturally\emph{\ traded} into the modular degrees of
freedom of \ $\mathfrak{M}_{g},_{N_{0}}$ representing punctures of the
underlying Riemann surface $((M;N_{0}),\mathcal{C})$. These punctures (which
vary holomorphically in $((M;N_{0}),\mathcal{C})$) are the points where we
have the coupling between $((M;N_{0}),\mathcal{C})$ and the $N_{0}(T)$
conformal degrees of freedom associated with the deficit angles $%
\{\varepsilon (k)\}$ (the curvature charges).\ As we shall see in the next
section, such a factorization of the conformal degrees of freedom into the
positional degrees of freedom of the interaction vertices and the degrees of
freedom associated with the corresponding curvature charges\ has important
consequences in understanding the dynamics of simplicial quantum gravity as
compared to its continuous field-theoretic counterpart.

\section{Trading conformal modes for moduli: The case of pure gravity}

In order to formalize a procedure for describing at the level of measures
the trading a (discretized) conformal degree of freedom into a moduli
parameter (a puncture), \ let us define the forgetful mapping 
\begin{equation}
\phi :\mathfrak{M}_{g},_{N_{0}}\times {R}_{+}^{N_{0}}\rightarrow \mathfrak{M}%
_{g},_{N_{0}}
\end{equation}
as the application which forgets the decorations on $\mathfrak{M}%
_{g},_{N_{0}}\times {R}_{+}^{N_{0}}$ provided by the perimeters $%
\{L(k)\}=\left( \frac{\sqrt{3}}{3}a\right) \{q(k)\}_{i=1}^{N_{0}(T)}$, (%
\emph{i.e.}, $\phi $ associates to $(\widetilde{\psi })$ the underlying
Riemann surface $((M;N_{0}),\mathcal{C})$ defined by (\ref{riemsurf})). Then
the map 
\begin{gather}
f_{T_{a}}:\Omega _{T_{a}}(\{q(k)\}_{k=1}^{N_{0}})\overset{h^{-1}}{%
\longrightarrow }\mathfrak{M}_{g},_{N_{0}}\times {R}_{+}^{N_{0}}\overset{\phi }{%
\rightarrow }\mathfrak{M}_{g},_{N_{0}}  \label{cellular} \\
(|P_{T_{L}}|\rightarrow {M)\longmapsto }((M;N_{0}),\mathcal{C},\psi
)\longmapsto ((M;N_{0}),\mathcal{C})  \notag
\end{gather}
defines an open cell in the moduli space $\mathfrak{M}_{g},_{N_{0}}$, labelled
by the given (generalized) dynamical triangulation $|T_{l=a}|\rightarrow
M\in \mathcal{DT}[\{q(k)\}_{k=1}^{N_{0}}]$, (recall that $\Omega
_{T_{a}}(\{q(k)\}_{k=1}^{N_{0}})$ is the rational cell associated with the
dynamical triangulation $|T_{l=a}|\rightarrow M$; see (\ref{omega})).\ \ As $%
|T_{l=a}|\rightarrow M$ varies in the discrete set $\mathcal{DT}%
[\{q(k)\}_{k=1}^{N_{0}}]$, we get in this way a cell decomposition of $\mathfrak{%
M}_{g},_{N_{0}}$ which is parametrized by $\mathcal{DT}[\{q(k)%
\}_{k=1}^{N_{0}}]$, \emph{i.e.}, 
\begin{equation}
\mathfrak{M}_{g},_{N_{0}}\left( \{q(k)\}_{k=1}^{N_{0}}\right)
=\bigsqcup_{T_{a}\in \mathcal{DT}[\{q(k)\}_{k=1}^{N_{0}}]}f_{T_{a}}\left(
\Omega _{T_{a}}(\{q(k)\}_{k=1}^{N_{0}})\right) .  \label{cell}
\end{equation}
Note that, as the notation suggests, such a cell decomposition depends on
the set of curvature assignments considered $\{q(k)\}_{k=1}^{N_{0}}$.
Distinct curvature assignments $\{\widehat{q(k)}\}_{k=1}^{N_{0}}$ gives rise
to possibly distinct cell decompositions of $\mathfrak{M}_{g},_{N_{0}}$ labelled
by the corresponding $\mathcal{DT}[\{\widehat{q(k)}\}_{k=1}^{N_{0}}]$.

It is worthwhile to note \ that the cell decomposition (\ref{cell})\ is the
combinatorial counterpart of the slice theorem (\ref{meslice}). There, we
identified a local slice for the action of \ $W(M)\ltimes Diff(M)$ with a
deformation of a Riemann surface (\emph{i.e.}, a point of $\mathfrak{M}_{g}$)
generated by a (holomorphic) quadratic differential, together with a family
of corresponding conformal factors on the fibers. Here, such a local slice
is combinatorially described by the set of triangulations $\mathcal{DT}%
[\{q(k)\}_{k=1}^{N_{0}}]\ $which through the corresponding meromorphic
quadratic differentials $((M;N_{0}),\mathcal{C},\psi )$ also generate
deformations of \ Riemann surfaces, (now represented by points of $\mathfrak{M}%
_{g},_{N_{0}}$), and a family of (discretized) conformal degrees of freedom
on the fibers provided by the curvature charges $\oint_{\partial (\rho
^{2}(k))}\sqrt{\psi (k)}$\ associated with $\psi $.

\bigskip

If \ the cell decomposition (\ref{cell})\ is the combinatorial counterpart
of the slice theorem (\ref{meslice}) then, the combinatorial aspects of the
anomalous scaling \ of the Weyl measure $D_{h}[v]$, (see also (\ref{dhoker}%
)), should be related to integration over the moduli space $\overline{\mathfrak{M%
}}_{g},_{N_{0}}$. In order to prove this latter statement,\ let us denote by 
$\omega _{WP}$ the Weil-Petersson $2$-form on $\overline{\mathfrak{M}}%
_{g},_{N_{0}}$ (see \emph{e.g.} (\ref{W-P}) in the appendix for a
definition). \ \ According to a recent result [11] by P. Zograf, one can
explicitly represent $\omega _{WP}$ in terms of the Witten-Kontsevich
intersection numbers. Since these latter can be expressed in terms of the
form (\ref{chern}) defined on $\Omega _{T_{a}}(\{q(k)\}_{k=1}^{N_{0}})$, it
follows that\ under the \ map $f_{T_{a}}$ defined by (\ref{cellular}), we
can pull back $\omega _{WP}$\ to a form $f_{T_{a}}^{\ast }(\omega _{WP})$ \
defined on the cellular orbicells $\Omega _{T_{a}}(\{q(k)\}_{k=1}^{N_{0}})$.
Explicitly, let us consider, for each partition $%
d_{1}+d_{2}+...+d_{N_{0}}=3g-3+N_{0}(T)$, of $\dim _{C}\overline{\mathfrak{M}}%
_{g},_{N_{0}}$\ \ the form 
\begin{equation}
\tau _{0}^{l_{0}}\tau _{1}^{l_{1}}...\tau
_{3g-3+N_{0}}^{l_{3g-3+N_{0}}}\doteq \omega _{1}^{d_{1}}\wedge \omega
_{2}^{d_{2}}\wedge ...\wedge \omega _{N_{0}}^{d_{N_{0}}},  \label{Kform}
\end{equation}
where each $\omega _{k}$ is the (combinatorial) Chern form (\ref{chern})
associated with the generic Regge polytope $(|P_{T_{L}}|\rightarrow {M)}$ in 
$\Omega _{T_{a}}(\{q(k)\}_{k=1}^{N_{0}})$, and where the notation $\tau
_{0}^{l_{0}}\tau _{1}^{l_{1}}...\tau _{3g-3+N_{0}}^{l_{3g-3+N_{0}}}$ keeps
track of the integers $\{l_{k}\geq 0\}$ enumerating how many of the summands 
$d_{i\text{ }}$are equal to $k$. The integral of $\tau _{0}^{l_{0}}\tau
_{1}^{l_{1}}...\tau _{3g-3+N_{0}}^{l_{3g-3+N_{0}}}$ over the cellular
decomposition of $\overline{\mathfrak{M}}_{g},_{N_{0}}$ provided by (\ref{cell})
defines the (combinatorial) Witten-Kontsevich intersection numbers $%
\left\langle \tau _{0}^{l_{0}}\tau _{1}^{l_{1}}...\tau
_{3g-3+N_{0}}^{l_{3g-3+N_{0}}}\right\rangle $ of $\overline{\mathfrak{M}}%
_{g},_{N_{0}}$, [8], \ [9],[21]. In terms of \ the family of forms\ (\ref
{Kform}) the Weil-Petersson volume form associated with the cellular
decomposition (\ref{cell}) can be written as [11] 
\begin{gather}
\frac{f_{T_{a}}^{\ast }(\omega _{WP})^{3g-3+N_{0}(T)}}{\left(
3g-3+N_{0}(T)\right) !}=  \label{zomeas} \\
\notag \\
=\sum_{\left| l\right| =3g-3+N_{0}(T)}\frac{(-1)^{g-1+N_{0}(T)+\left\|
l\right\| }}{\prod_{i=2}^{3g-2+N_{0}(T)}l_{i}!\left( (i-1)!\right) ^{l_{i}}}%
\tau _{0}^{N_{0}(T)}\tau _{2}^{l_{2}}...\tau _{3g-2+N_{0}}^{l_{3g-2+N_{0}}},
\notag
\end{gather}
where $l\doteq (l_{2},l_{3},...,l_{3g-2+N_{0}(T)})$ is a multi-index and
where 
\begin{equation}
\begin{tabular}{ccc}
$|l|$ & $\doteq $ & $\sum_{k=2}^{3g-2+N_{0}(T)}(k-1)l_{k}$ \\ 
&  &  \\ 
$\left\| l\right\| $ & $\doteq $ & $\sum_{k=2}^{3g-2+N_{0}(T)}l_{k}$%
\end{tabular}
.
\end{equation}

\noindent With these preliminary remarks along the way, let us consider the
Weil-Petersson volume $VOL\left( \overline{\mathfrak{M}}_{g},_{N_{0}}\right) $\
of the (compactified) moduli space $\overline{\mathfrak{M}}_{g},_{N_{0}}$, 
\begin{gather}
VOL\left( \overline{\mathfrak{M}}_{g},_{N_{0}}\right) =\frac{1}{N_{0}!}\int_{%
\overline{\mathfrak{M}}_{g},_{N_{0}}}\frac{\omega _{WP}{}^{3g-3+N_{0}(T)}}{%
\left( 3g-3+N_{0}(T)\right) !}=  \label{wpvol} \\
\notag \\
=\frac{1}{N_{0}!}\sum_{T\in \mathcal{DT}[\{q(i)\}_{i=1}^{N_{0}}]}%
\int_{f_{T_{a}}\left( \Omega _{T_{a}}(\{q(k)\}_{k=1}^{N_{0}})\right) }\frac{%
\omega _{WP}{}^{3g-3+N_{0}(T)}}{\left( 3g-3+N_{0}(T)\right) !}  \notag
\end{gather}
where we have exploited the cell decomposition (\ref{cell}) of $\overline{%
\mathfrak{M}}_{g},_{N_{0}}$, and where we have divided by $N_{0}(T)!$ in order
to factor out the labelling of the $N_{0}(T)$ punctures. \\ 
By pulling $%
\omega _{WP}$ back to the orbicells $\Omega _{T_{a}}(\{q(k)\}_{k=1}^{N_{0}})$
by means of the map $f_{T_{a}}$ and by taking into account that each $\Omega
_{T_{a}}(\{q(k)\}_{k=1}^{N_{0}})$ is a (smooth) orbifold acted upon by the
automorphism group $Aut_{\partial }(P_{T_{a}})$, we can explicitly write (%
\ref{wpvol}) as an orbifold integration,(the orbifold integration over
moduli space and consistent with the orientation of $\overline{\mathfrak{M}}%
_{g},_{N_{0}}$ is defined in [28], Th. 3.2.1), according to

\begin{gather}
VOL\left( \overline{\mathfrak{M}}_{g},_{N_{0}}\right) =\frac{1}{N_{0}!}\int_{%
\overline{\mathfrak{M}}_{g},_{N_{0}}}\frac{\omega _{WP}{}^{3g-3+N_{0}(T)}}{%
\left( 3g-3+N_{0}(T)\right) !}=  \label{orbin} \\
\notag \\
=\frac{1}{N_{0}!}\sum_{T\in \mathcal{DT}[\{q(i)\}_{i=1}^{N_{0}}]}\frac{1}{%
|Aut_{\partial }(P_{T_{a}})|}\int_{_{\Omega
_{T_{a}}(\{q(k)\}_{k=1}^{N_{0}})}}\frac{f_{T_{a}}^{\ast }(\omega
_{WP})^{3g-3+N_{0}(T)}}{\left( 3g-3+N_{0}(T)\right) !},  \notag
\end{gather}
where the integration measure is defined by (\ref{zomeas}), and \ where the
summation is over all distinct dynamical triangulations with given unlabeled
curvature assignments weighted by the order $|Aut_{\partial }(P_{T})|$ of
the automorphisms group of the corresponding dual polytope. We need the
following technical

\begin{lemma}
For any orbicell $\Omega _{T_{a}}(\{q(k)\}_{k=1}^{N_{0}})$ we have\ 
\begin{equation}
\lim_{N_{0}\rightarrow \infty }\int_{_{\Omega
_{T_{a}}(\{q(k)\}_{k=1}^{N_{0}})}}\frac{f_{T_{a}}^{\ast }(\omega
_{WP})^{3g-3+N_{0}(T)}}{\left( 3g-3+N_{0}(T)\right) !}\longrightarrow \delta
_{\widetilde{T}_{a}},
\end{equation}
where the limit is in the sense of distributions, and where $\delta _{T_{a}}$
denotes the Dirac measure concentrated on a dynamical triangulation $|%
\widetilde{T}_{l=a}|\rightarrow M\in \mathcal{DT}[\{q(i)\}_{i=1}^{\infty }]$
with the essential incidence of \ the given $T_{a}\in \mathcal{DT}%
[\{q(k)\}_{k=1}^{N_{0}}]$,\emph{\ i.e.}, with 
\begin{equation}
\{q(i)\}_{i=1}^{\infty }=\left\{
\{q(k)\}_{k=1}^{N_{0}},\{q(h)\}_{N_{0}+1}^{\infty }=6\right\} .
\label{hexassign}
\end{equation}
\end{lemma}

In order to prove this statement, let us recall that according to a result
of C.McMullen [29] \ the Weil-Petersson form $\omega _{WP}$ can be written
as $d\Theta $ for some bounded $1$-form $\Theta $. For instance, one can use
the representation 
\begin{equation}
\Theta =-i\beta _{M},\;\;d\Theta =\omega _{WP},
\end{equation}
where $\beta _{M}$ is the Bers embedding

\begin{equation}
\beta _{M}:\mathfrak{T}_{g,N_{0}}(M)\rightarrow Q_{N_{0}}(M)\simeq T_{M}^{\ast }%
\mathfrak{T}_{g,N_{0}}(M)\text{,}  \label{bers}
\end{equation}
\ and where $Q_{N_{0}}(M)$ denotes the space of (holomorphic) quadratic
differentials (see the appendix for a characterization of $\beta _{M}$).
This implies that for any compact complex submanifold $V^{2m}\subset \mathfrak{T}%
_{g,N_{0}}(M)$, one has 
\begin{equation}
Vol_{WP}(V^{2m})=\int_{V^{2m}}\omega _{WP}^{m}=\int_{\partial V}\Theta
\wedge \omega _{WP}^{m-1},  \label{surf}
\end{equation}
where the $2m-1$ form $\Theta \wedge \omega _{WP}^{m-1}$ is bounded. The
cellular decomposition $f_{T_{a}}$ defined by (\ref{cellular}) may be
thought of as representing the combinatorial counterpart of the Bers
embedding,\ \ and, if $V^{2m}$ is the image under the map (\ref{cellular})
of \ the closure $\overline{\Omega }_{T_{a}}$ of \ $\Omega
_{T_{a}}(\{q(k)\}_{k=1}^{N_{0}})$, we can write 
\begin{gather}
\int_{\overline{\Omega }_{T_{a}}}\frac{f_{T_{a}}^{\ast }(\omega
_{WP})^{3g-3+N_{0}(T)}}{\left( 3g-3+N_{0}(T)\right) !}=\int_{\partial
\lbrack \overline{\Omega }_{T_{a}}]}\frac{f_{T_{a}}^{\ast }\Theta \wedge
f_{T_{a}}^{\ast }(\omega _{WP})^{3g-4+N_{0}(T)}}{\left( 3g-3+N_{0}(T)\right)
!}=  \label{leray} \\
=\delta _{\partial \overline{\Omega }_{T_{a}}},  \notag
\end{gather}
where we have considered 
\begin{equation}
f_{T_{a}}^{\ast }(\omega _{WP})^{3g-4+N_{0}(T)}
\end{equation}
as a Leray form based on the hypersurface $\partial \overline{\Omega }%
_{T_{a}}$\ and have interpreted the surface integral appearing in (\ref
{leray}) as the definition (with respect to the Weil-Petersson volume) of
the Dirac measure $\delta _{\partial \overline{\Omega }_{T_{a}}}$ based on
the hypersurface $\partial \overline{\Omega }_{T_{a}}$. With such a
characterization of the surface Dirac measure associate with $%
f_{T_{a}}^{\ast }(\omega _{WP})$, \ the lemma follows from (\ref{leray}) if
we can show that as $N_{0}\rightarrow \infty $, $\partial \overline{\Omega }%
_{T_{a}}$ converges suitably to a dynamical triangulation $|\widetilde{T}%
_{l=a}|\rightarrow M\in \mathcal{DT}[\{q(i)\}_{i=1}^{\infty }]$. To this
end, let us fix some $N$ large enough and consider for $N_{0}<<N$ \ the
convex subset $\Delta _{T_{a}}(\{q(k)\}_{k=1}^{N_{0}})$\ of $\mathbb{R}%
^{2N+6g-6}$ associated with the orbicell $\Omega
_{T_{a}}(\{q(k)\}_{k=1}^{N_{0}})$. Let $|T_{l}|\rightarrow M$ a
triangulation in $\Delta _{T_{a}}(\{q(k)\}_{k=1}^{N_{0}})$ distinct from the
given dynamical triangulation $|T_{l=a}|\rightarrow M$. By iterating an hex
refinement on $|T_{l}|\rightarrow M$ we can always assume that for $%
N\rightarrow \infty $, the triangulation $|T_{l}|\rightarrow M$ converges
(in the standard topology of $\mathbb{R}^{2N+6g-6}$) to a dynamical
triangulation $|\widetilde{T}_{l=a}|\rightarrow M$ with curvature
assignments given by (\ref{hexassign}). Under the same sequence of hex
refinement also $|T_{l=a}|\rightarrow M$ converges to $|\widetilde{T}%
_{l=a}|\rightarrow M$. Since $|T_{l}|\rightarrow M$ is a generic
triangulation of $\Delta _{T_{a}}(\{q(k)\}_{k=1}^{N_{0}})$, it follows that $%
\Delta _{T_{a}}(\{q(k)\}_{k=1}^{N_{0}})$ shrinks, as $N\rightarrow \infty $,
to $|\widetilde{T}_{l=a}|\rightarrow M$. All triangulations of $\Delta
_{T_{a}}(\{q(k)\}_{k=1}^{N_{0}})$ have the same automorphism group $%
Aut_{\partial }(P_{T_{a}})$\ of $|T_{l=a}|\rightarrow M$, and since the hex
refinement does not alter $Aut_{\partial }(P_{T_{a}})$, we get that in $%
\mathbb{R}^{2N+6g-6}$ 
\begin{equation}
\Omega _{T_{a}}(\{q(k)\}_{k=1}^{N_{0}})\longrightarrow \left( |\widetilde{T}%
_{l=a}|\rightarrow M\right) ,
\end{equation}
as $N\rightarrow \infty $. Going to the closure\ $\partial \overline{\Omega }%
_{T_{a}}\longrightarrow (|\widetilde{T}_{l=a}|\rightarrow M$ ) and $\delta
_{_{\partial \overline{\Omega }_{T_{a}}}}$ converges (weakly) to $\delta _{%
\widetilde{T}_{a}}$ as stated.

\bigskip

By introducing a smooth, compactly supported approximation to the
characteristic function of \ $\overline{\Omega }_{T_{a}}$\ we get from the
above lemma an immediate but important consequence

\begin{proposition}
In the large $N_{0}$ limit we have the asymptotic relation 
\begin{equation}
VOL\left( \overline{\mathfrak{M}}_{g},_{N_{0}}\right) \approx \frac{1}{N_{0}!}%
\sum_{T\in \mathcal{DT}[\{q(i)\}_{i=1}^{N_{0}}]}\frac{1}{|Aut_{\partial
}(P_{T_{a}})|}.  \label{asymp}
\end{equation}
\end{proposition}

Note that 
\begin{equation}
Card\left[ \mathcal{DT}[\{q(i)\}_{i=1}^{N_{0}}]\right] \doteq \frac{1}{N_{0}!%
}\sum_{T\in \mathcal{DT}[\{q(i)\}_{i=1}^{N_{0}}]}\frac{1}{|Aut_{\partial
}(P_{T_{a}})|}
\end{equation}
provides the number of distinct dynamical triangulations with given
(unlabeled) curvature assignments $\{q(i)\}_{i=1}^{N_{0}}$\ over the $%
N_{0}(T)$ vertices $[\{\sigma ^{0}(i)\}_{i=1}^{N_{0}}]$. Such a counting
function is related to the enumeration of all distinct (generalized)
triangulations with a given number $N_{0}$ of unlabeled vertices by 
\begin{equation}
Card\left[ \mathcal{DT}\left( N_{0}\right) \right] =\sum_{\{q(i)%
\}_{i=1}^{N_{0}}}Card\left[ \mathcal{DT}[\{q(i)\}_{i=1}^{N_{0}}]\right] 
\label{card}
\end{equation}
where the summation $\sum_{\{q(i)\}_{i=1}^{N_{0}}}$ is over all possible
curvature assignments on the $N_{0}$ (unlabeled) vertices. In simplicial
quantum gravity $Card\left[ \mathcal{DT}\left( N_{0}\right) \right] $ is a
basic quantity providing the canonical entropy function associated with the
set of triangulations considered [5]. \ Since $VOL\left( \overline{\mathfrak{M}}%
_{g},_{N_{0}}\right) $ does not depend on the particular cellular
decomposition defined by the given choice of the curvature assignments $%
\{q(i)\}_{i=1}^{N_{0}}$, according to (\ref{asymp}) and (\ref{card}) we get
the asymptotic relation 
\begin{equation}
Card\left[ \mathcal{DT}\left( N_{0}\right) \right] \approx VOL\left( 
\overline{\mathfrak{M}}_{g},_{N_{0}}\right) Card\left[ \{q(i)\}_{i=1}^{N_{0}}%
\right] ,  \label{factor}
\end{equation}
where $Card\left[ \{q(i)\}_{i=1}^{N_{0}}\right] $ denotes the number of
possible curvature assignments over the $N_{0}$ (unlabeled) vertices $%
[\{\sigma ^{0}(i)\}_{i=1}^{N_{0}}]$.\ \ (For caveats concerning the meaning
of this formula\ as compared to some of our previous work, see Appendix II).

\bigskip

At this stage, it is important to remark that we have two independent
asymptotic evaluations of \ $Card\left[ \mathcal{DT}\left( N_{0}\right) %
\right] $ and of $VOL\left( \overline{\mathfrak{M}}_{g},_{N_{0}}\right) $. Thus,
the relations (\ref{asymp}) and (\ref{factor}) provide a non trivial
geometrical interpretation of the canonical entropy function, $Card\left[ 
\mathcal{DT}\left( N_{0}\right) \right] $ which can be profitably confronted
with the field-theoretic measure. The large $N_{0}$ asymptotics of $%
Vol_{W-P}(\overline{\mathfrak{M}}_{g},_{N_{0}})$ has been discussed by Manin and
Zograf [10], [11]. They obtained 
\begin{gather}
Vol_{W-P}(\overline{\mathfrak{M}}_{g},_{N_{0}})=\pi ^{2(3g-3+N_{0})}\times
\label{manzog} \\
\times (N_{0}+1)^{\frac{5g-7}{2}}C^{-N_{0}}\left( B_{g}+\sum_{k=1}^{\infty }%
\frac{B_{g,k}}{(N_{0}+1)^{k}}\right) ,  \notag
\end{gather}
where $C=-\frac{1}{2}j_{0}\frac{d}{dz}J_{0}(z)|_{z=j_{0}}$, ($J_{0}(z)$\ the
Bessel function, $j_{0}$ its first positive zero); (note that $C\simeq
0.625....$). The genus dependent parameters $B_{g}$\ are explicitly given
[10] by 
\begin{equation}
\left\{ 
\begin{tabular}{ll}
$B_{0}=\frac{1}{A^{1/2}\Gamma (-\frac{1}{2})C^{1/2}},$ & $B_{1}=\frac{1}{48}%
, $ \\ 
$B_{g}=\frac{A^{\frac{g-1}{2}}}{2^{2g-2}(3g-3)!\Gamma (\frac{5g-5}{2})C^{%
\frac{5g-5}{2}}}\left\langle \tau _{2}^{3g-3}\right\rangle ,$ & $g\geq 2$%
\end{tabular}
\right.
\end{equation}
where $A\doteq -j_{0}^{-1}J_{0}^{\prime }(j_{0})$, and $\left\langle \tau
_{2}^{3g-3}\right\rangle $ is a Kontsevich-Witten intersection number, (the
coefficients $B_{g,k}$ can be computed similarly-see [10] for details).
Thus, from (\ref{asymp}) we get that in the large $N_{0}$ limit 
\begin{equation}
Card\left[ \mathcal{DT}[\{q(i)\}_{i=1}^{N_{0}}]\right] \approx B_{g}\pi
^{2(3g-3+N_{0})}N_{0}{}^{\frac{5g-7}{2}}C^{-N_{0}}\left( 1+O(\frac{1}{N_{0}}%
)\right) ,  \label{berno}
\end{equation}
which does not depend on the actual distribution of the curvature
assignments $\{q(i)\}_{i=1}^{N_{0}}$, (this is not surprising since in the
large $N_{0}$ limit, the average number of triangles incident on a generic
vertex is $q(i)=6$, see (\ref{vincolo})). It is important to stress that $%
Card\left[ \mathcal{DT}[\{q(i)\}_{i=1}^{N_{0}}]\right] $ is the
combinatorial counterpart of the local slice in $Riem(M)/Diff(M)$ defined by
the metrics with constant curvature. In this connection, note that (\ref
{berno})\ \ shows that it is precisely this part of the full triangulation
counting (\ref{factor}) that contributes the genus-$g$ pure gravity critical
exponent

\begin{equation}
\gamma _{g}=\frac{5g-1}{2}.
\end{equation}
The relations (\ref{manzog}) \ and (\ref{berno}) also show that $\gamma _{g}$
has a modular origin, arising from the cardinality of the cell decomposition
of $\overline{\mathfrak{M}}_{g},_{N_{0}}$. To go deeper into such an issue, let
us recall that the Weil-Petersson volume of the moduli space $\overline{
\mathfrak{M}}_{g},_{N_{0}}$ for any fixed value of \ $N_{0}$\ is such that 
\begin{equation}
C_{1}^{g}(2g)!\leq Vol_{W-P}(\overline{\mathfrak{M}}_{g},_{N_{0}})\leq
C_{2}^{g}(2g)!,
\end{equation}
where the constants $0<C_{1}<C_{2}$ are independent of $N_{0}$ (see [30], [31]).
Thus the modular volume $Vol_{W-P}(\overline{\mathfrak{M}}_{g})$, (assuming $
g\geq 2$, or inserting at least $3$ stabilizing punctures), does not
contribute to the anomalous scaling term $N_{0}(T)^{\frac{5g-7}{2}}$ in the
asymptotics (\ref{berno}) for $Card\left[ \mathcal{DT}[\{q(i)
\}_{i=1}^{N_{0}}]\right] $. Such a scaling is generated only by the modular
parameters associated with the location of vertices, (this does not
surprise, since $\gamma _{g}$ enters directly topological triangulations
counting regardless of any metric structure the triangulation can carry).
Recall that there is an explicit connection (\ref{tzeta}) between the
variables $\{\zeta (k)\}$ uniformizing the polygonal cells of \ $
|P_{T_{a}}|\rightarrow {M}$ and the variables $\{t_{k}\}$ uniformizing the
corresponding vertex stars in $|T_{a}|\rightarrow {M}$. We have also an
explicit expression \ $v\doteq u-\sum_{k=1}^{N_{0}(T)}\left( \frac{%
\varepsilon (k)}{2\pi }\right) \ln \left| t-t_{k}\right| $\ (see\ (\ref
{conform}) ) for the conformal factor defining the metric of the
triangulation $|T_{a}|\rightarrow {M}$ in terms of the underlying
conformally flat structure (see (\ref{cmetr})). In such a framework we are
fully entitled to interpret the above remark, which sees the scaling as
associated with the modular parameters \ describing the punctures, as the
counterpart in DT theory of the anomalous scaling of the formal Weyl measure 
$D_{h}[v]$.

As mentioned, the large $N_{0}(T)$ asymptotics of the full triangulation
counting $Card\left[ \mathcal{DT}[N_{0}]\right] $\ can be obtained from
purely combinatorial (and matrix theory) arguments [5], [32] to the effect
that 
\begin{equation}
Card\left[ \mathcal{DT}[N_{0}]\right] \sim \frac{16c_{g}}{3\sqrt{2\pi }}%
\cdot e^{\mu _{0}N_{0}(T)}N_{0}(T)^{\frac{5g-7}{2}}\left( 1+O(\frac{1}{N_{0}}%
)\right) ,  \label{oldentr}
\end{equation}
where $c_{g}$ is a numerical constant depending only on the genus $g$, and $%
e^{\mu _{0}}=(108\sqrt{3})$ is a (non-universal) parameter depending on the
set of triangulations considered (here the generalized triangulations,
barycentrically dual to trivalent graphs; in the case of regular
triangulations in place of $108\sqrt{3}$ we would get $e^{\mu _{0}}=(\frac{%
4^{4}}{3^{3}})$). As already stressed, such a parameter does not play any
relevant role in 2D quantum gravity. Through a comparison of the asymptotics
(\ref{berno}) and (\ref{oldentr}), the relation (\ref{factor}) immediately
yields that 
\begin{equation}
Card\left[ \{q(i)\}_{i=1}^{N_{0}}\right] \sim \frac{c_{g}}{\pi ^{6g-6}B_{g}}%
\left[ \frac{Ce^{\mu _{0}}}{\pi ^{2}}\right] ^{N_{0}(T)},
\end{equation}
where corresponding to the given values of the parameters $C$ and $e^{\mu
_{0}}$ (see (\ref{manzog}) and (\ref{oldentr})) we have $(Ce^{\mu _{0}}/\pi
^{2})\sim 11.846$. \ Thus, as $N_{0}\rightarrow \infty $, the cardinality $%
Card\left[ \{q(i)\}_{i=1}^{N_{0}}\right] $ of the set of possible curvature
assignments grows exponentially, and only provides a non-universal
contribution to the whole triangulation counting. Summarizing the results
obtained so far we can state the following

\begin{proposition}
The anomalous scaling behavior of the canonical entropy function $Card\left[ 
\mathcal{DT}[N_{0}]\right] $ comes only from $Card\left[ \mathcal{DT}%
[\{q(i)\}_{i=1}^{N_{0}}]\right] $. In particular, up to an overall
normalization, the combinatorial counterpart of the anomalous scaling of the
field theoretic measure 
\begin{equation}
Z_{(n)}^{\pm }(h)D_{h}[v]\prod_{\beta =1}^{3g-3}dm_{\beta }d\overline{m}%
_{\beta },
\end{equation}
is provided, in the case of pure gravity, by the large $N_{0}$ asymptotics
of 
\begin{equation}
Vol_{W-P}(\overline{\mathfrak{M}}_{g},_{N_{0}})\exp \left[ -vN_{0}(T)\right]
\end{equation}
where\ $\nu \doteq 2\ln \frac{\pi }{C}$.
\end{proposition}

The overall normalization referred to in the above statement simply reflects
the presence of the leading exponential contributions to triangulations
counting associated both to $Card\left[ \{q(i)\}_{i=1}^{N_{0}}\right] $ and
to the term \ $\pi ^{2N_{0}}C^{-N_{0}}$ \ in \ (\ref{berno}). Such terms
affect the normalization \ $Z_{g}[1]$ in the continuum expression (\ref
{scaling1}) for the scaling, with the surface area, of the partition
function of 2D gravity. The exponential term $\exp \left[ -vN_{0}(T)\right] $
takes care of such normalization and isolates the polynomial subleading
asymptotics of $Vol_{W-P}(\overline{\mathfrak{M}}_{g},_{N_{0}})$ which is
responsible, according to (\ref{asymp}), of the anomalous scaling in
dynamical triangulation theory.

\bigskip

\section{Concluding remarks: matter fields}

From a moduli theory point of view, \ the above analysis\ explains the
origin of the critical exponents in the canonical entropy function (\ref
{oldentr}) for pure gravity. According to (\ref{factor}) and to the
Manin-Zograf asymptotics, the critical exponents are related to the
subleading polynomial asymptotics of the volume of the moduli space $%
\overline{\mathfrak{M}}_{g},_{N_{0}}$. In order to extend the factorization (\ref
{factor}) to the case of gravity coupled to matter, we need to discuss how
the discretization of the matter fields interacts with such factorization.
If we parametrize the set of possible conformal fields $\{\Phi \}$ over $M$
in terms of \ a smooth projective manifold (actually a variety) of dimension 
$r$, then a basic ingredient of the quantum theory describing the
interaction of $\{\Phi \}$ with $2$D gravity is the path integral over the
space of maps $(M,\mathcal{C})\rightarrow X$ , between a marked Riemann
surface and the target $X$. Such path integrals tend to localize around the
space of holomorphic maps which thus provides a sort of discretization of
the matter functional integral, (see \emph{e.g.}, [12]). In a simplicial
quantum gravity setting a distribution of matter fields $\{\Phi \}$ over a
dynamical triangulation \ $|T_{a}|\rightarrow {M}$ can be described by a map 
\begin{gather}
\overline{\mathfrak{M}}_{g},_{N_{0}}\rightarrow RGB_{g,N_{0}}^{met}\times X
\label{rem1} \\
((M;N_{0}),\mathcal{C})\longmapsto (\Gamma ,\Phi _{(k)})  \notag
\end{gather}
where \ \ $((M;N_{0}),\mathcal{C})$ is the punctured Riemann surface
associated with $|T_{a}|\rightarrow {M}$, $\Gamma $ is the corresponding
ribbon graph in $RGB_{g,N_{0}}^{met}$, and $\Phi _{(k)}$ is a locally
constant map describing conformal fields attached to the polygonal faces $%
\{\rho ^{2}(k)\}$ of \ the corresponding dual polytope $|P_{Ta}|\rightarrow {%
M}$. According to the bijection (\ref{bijec}), we can write (\ref{rem1})
equivalently as 
\begin{equation}
\overline{\mathfrak{M}}_{g},_{N_{0}}\rightarrow \left( \overline{\mathfrak{M}}%
_{g},_{N_{0}}\times X\right) \times \mathbb{R}_{+}^{N}.
\end{equation}
Since $\overline{\mathfrak{M}}_{g},_{N_{0}}\times X$ is isomorphic to the space $%
\overline{\mathfrak{M}}_{g},_{N_{0}}(X,0)$\ of stable constant maps from $
((M;N_{0}),\mathcal{C})$ to\ \ $X$ , we have a natural candidate which
should play in the matter setting the same role of $\overline{\mathfrak{M}}
_{g},_{N_{0}}$ in the pure gravity case. More generally, it is technically
convenient to introduce the space $\overline{\mathfrak{M}}_{g},_{N_{0}}(X,\beta
) $\ of stable maps from $((M;N_{0}),\mathcal{C})$ to\ \ $X$ \ representing
the homology class $\beta \in H_{2}(X,\mathbb{Z})$, (for instance, in the
case of the Polyakov action, the map can be thought of \ as providing the
embedding of $((M;N_{0}),\mathcal{C})$ in the target (Euclidean) space).
According to these remarks it is rather natural to conjecture that, for
dynamical triangulations, the measure describing the statistical distribution of
conformal matter interacting with 2D gravity is provided by
a (generalized) Weil-Petersson volume of $\overline{\mathfrak{M}}
_{g},_{N_{0}}(X,\beta )$. This observation calls into play the (descendent)\
Gromov-Witten invariants of the manifold $X$ \ (in order to provide the
analogous of \ (\ref{zomeas}))\ \ and to a large extent can be discussed
along the same lines of the pure gravity case. As already mentioned in the
introduction, such a connection is in line with the fact that many relevant
properties of $\overline{\mathfrak{M}}_{g},_{N_{0}}(X,\beta )$ are governed, via
its intersection theory, by matrix models. This is a consequence of the
localization of path integrals over maps (conformal matter) which computes
such path integrals as sums, labelled by ribbon graphs, of finite
dimensional integrals over space of stable holomorphic maps. We thus expect
that the techniques related to the study of \ the properties of\ $\overline{
\mathfrak{M}}_{g},_{N_{0}}(X,\beta )$, on which there has been much progress
recently [12], \ will led to a deeper understanding of \ the anomalous
scaling of conformal matter interacting with 2D gravity (\emph{i.e.},
non-critical strings).
\vfill\eject
\section{\protect\bigskip Appendix I : Teichm\"{u}ller theory of pointed
Riemann surfaces}

In this section we summarize, for the reader's convenience, basic facts
about Teichm\"{u}ller space theory. For details and proofs we refer to
[33].\ \ If $S_{2}(M)$ denotes the space of symmetric bilinear forms on $M$,
let us\ consider the set of all Riemannian metrics on $(M;N_{0})$, \emph{i.e.
} 
\begin{equation}
Riem(M;N_{0})\doteq \left\{ g\in S_{2}(M)|\;g(x)(u,u)>0\;if\;u\neq 0\right\}
,
\end{equation}
and let 
\begin{equation}
Riem_{-1}(M;N_{0})\hookrightarrow Riem(M;N_{0})
\end{equation}
be the set of metrics of constant curvature $-1$, describing the hyperbolic
structures on $(M;N_{0})$. If $Diff_{+}(M)$ is the group of all orientation
preserving diffeomorphisms then\ 
$$Diff_{+}(M;N_{0})=$$
\begin{equation}
=\left\{ \psi \in Diff_{+}(M):\psi
\;preserves\;setwise\;\{\sigma ^{0}(i)\}_{i=1}^{N_{0}(T)}\right\}
\end{equation}
acts by pull-back on the metrics in $Riem_{-1}(M;N_{0})$. Let $%
Diff_{0}(M;N_{0})$ be the subgroup consisting of diffeomorphisms which when
restricted to $(M;N_{0})$ are isotopic to the identity, then the
Teichm\"{u}ller space $\mathfrak{T}_{g,N_{0}}(M)$ associated with the genus $g$
surface with $N_{0}(T)$ punctures $M$ is defined by 
\begin{equation}
\mathfrak{T}_{g,N_{0}}(M)=\frac{Riem_{-1}(M;N_{0})}{Diff_{0}(M;N_{0})}.
\end{equation}
Recall that a Riemann surface is a complex analytic structure on $M$
consisting of an atlas of charts $\{U_{\alpha },\zeta _{\alpha }\}$ where \ $%
\{U_{\alpha }\}$ is a covering of $M$ by open sets, $\zeta _{\alpha
}:U_{\alpha }\rightarrow \mathbb{C}$\ is a homeomorphism and if \ $U_{\alpha
}\cap U_{\beta }\neq \emptyset $, then $\zeta _{\alpha }\circ \zeta _{\beta
}^{-1}:\zeta _{\beta }(U_{\alpha }\cap U_{\beta })\rightarrow \mathbb{C}$\
is complex analytic. The local maps $\zeta _{\alpha }:U_{\alpha }\rightarrow 
\mathbb{C}$ are the uniformizing parameters of the surface. From such a
complex function theory perspective, $\mathfrak{T}_{g,N_{0}}(M)$ is defined by
fixing a reference complex structure $\mathcal{C}_{0}$ on $(M;N_{0})$ (a
marking) and considering the set of equivalence classes of complex
structures $(\mathcal{C},f)$\ where $f:\mathcal{C}_{0}\rightarrow \mathcal{C}
$ is an orientation preserving quasi-conformal map, and where any two pairs
of complex structures $(\mathcal{C}_{1},f_{1})$ and $(\mathcal{C}_{2},f_{2})$
are considered equivalent if $h\circ f_{1}$ is homotopic to $f_{2}$ via a
conformal map\ $h:\mathcal{C}_{1}\rightarrow \mathcal{C}_{2}$. Let us assume
that the reference complex structure $\mathcal{C}_{0}$ admits an
antiholomorophic reflection $j:\mathcal{C}_{0}\rightarrow \mathcal{C}_{0}$.
Since any orientation reversing diffeomorphism $\widetilde{\varphi }$ can be
written as $\varphi \circ j$ for some orientation preserving diffeomorphism $%
\varphi $,\ the (extended) mapping class group $\mathfrak{Map}(M;N_{0})\doteq
Diff(M;N_{0})/Diff_{0}(M;N_{0})$ \ acts naturally on $\mathfrak{T}_{g,N_{0}}(M)$
according to 
\begin{gather}
\mathfrak{Map}(M;N_{0})\times \mathfrak{T}_{g,N_{0}}(M)\longrightarrow \mathfrak{T}%
_{g,N_{0}}(M) \\
\left\{ 
\begin{tabular}{l}
$(\varphi ,(\mathcal{C},f))\longmapsto (\mathcal{C},f\circ \varphi ^{-1}),$\\ 
$(\widetilde{\varphi },(\mathcal{C},f))\longmapsto (\mathcal{C}^{\ast
},f\circ j\circ \varphi ^{-1})$ 
\end{tabular}
\right. ,   \notag
\end{gather}
where 
$$\left\{ 
\begin{tabular}{l}
$\varphi \in Diff_{+}(M;N_{0})$ \\
$\widetilde{\varphi}\in Diff(M;N_{0})-Diff_{+}(M;N_{0})$
\end{tabular}\right. ,$$
and where the conjugate surface $\mathcal{C}^{\ast }$ is the Riemann surface
locally described by the complex conjugate coordinate charts associated with 
$\mathcal{C}$. It follows that the Teichm\"{u}ller space $\mathfrak{T}%
_{g,N_{0}}(M)$ can be also seen as the universal cover of the moduli space $%
\mathfrak{M}_{g},_{N_{0}}$ of genus $g$ Riemann surfaces with $N_{0}(T)$\
punctures defined by 
\begin{equation}
\mathfrak{M}_{g},_{N_{0}}=\frac{\mathfrak{T}_{g,N_{0}}(M)}{\mathfrak{Map}(M;N_{0})}
\end{equation}
It is well-known that $\mathfrak{M}_{g},_{N_{0}}$ is a connected orbifold space
of complex dimension \ $3g-3+N_{0}(T)$ and that, although in general non
complete, it admits a \ stable curve compactification (Deligne-Mumford) $%
\overline{\mathfrak{M}}_{g},_{N_{0}}$. The orbifold $\overline{\mathfrak{M}}%
_{g},_{N_{0}}$ is endowed with $N_{0}(T)$ natural line bundles $\mathcal{L}%
_{i}$ ( the cotangent space to $M$ at the $i$-th marked point), and with a
natural rank $g$ vector bundle $\mathbb{E}$, the Hodge bundle whose fibers
are top wedge powers of Abelian differentials on the surface.

For \ a genus $g$ Riemann surfaces with $N_{0}(T)>3$\ punctures the complex
vector space $Q_{N_{0}}(M)$ of (holomorphic) quadratic differentials is
defined by tensor fields $\psi $ described, in a locally uniformizing
complex coordinate chart $(U,\zeta )$, by a holomorphic function $\mu
:U\rightarrow \mathbb{C}$ such that $\psi =\mu (\zeta )d\zeta \otimes d\zeta 
$. \ Away from the discrete set of the zeros of \ $\psi $, we can locally
choose a canonical conformal coordinate $z_{(\psi )}$ (unique up to $%
z_{(\psi )}\mapsto \pm z_{(\psi )}+const$) by integrating the holomorphic $1$%
-form $\sqrt{\psi }$, \emph{i.e}., \ 
\begin{equation}
z_{(\psi )}=\int^{\zeta }\sqrt{\mu (\zeta ^{\prime })d\zeta ^{\prime
}\otimes d\zeta ^{\prime }},  \label{holqua}
\end{equation}
so that $\psi =dz_{(\psi )}\otimes dz_{(\psi )}$. 
\\ If we endow $Q_{N_{0}}(M)$ with the $L^{1}$-(Teichm\"{u}ller) norm 
\begin{equation}
||\psi ||\doteq \int_{M}|\psi |,  \label{teichnorm}
\end{equation}
\ then the Banach space of integrable quadratic differentials on $M$, $%
Q_{N_{0}}(M)\doteq \{\psi |,\;||\psi ||<+\infty \}$,\ is non-empty and
consists of meromorphic quadratic differentials whose only singularities are
(at worst) simple poles at the $N_{0}$ distinguished points of $(M,N_{0})$. $%
Q_{N_{0}}(M)$ is finite dimensional and, according to the Riemann-Roch
theorem, it has complex dimension $\dim _{\mathbb{C}%
}\,Q_{N_{0}}(M)=3g-3+N_{0}(T)$.

From the viewpoint of Riemannian geometry, a quadratic differential is
basically a transverse-traceless two tensor deforming a Riemannian structure
to a nearby inequivalent Riemannian structure. Thus a quadratic differential 
$\psi =\mu (\zeta )d\zeta \otimes d\zeta $ also encodes information on
possible deformations of the given complex structure. Explicitly, by
performing an affine transformation with constant dilatation $K>1$, one
defines a new uniformizing variable $z_{(\psi )}^{\prime }$ associated with $%
\psi =\mu (\zeta )d\zeta \otimes d\zeta $ by deforming the variable $%
z_{(\psi )}$ defined by (\ref{holqua}) according to 
\begin{equation}
z_{(\psi )}^{\prime }=K\func{Re}(z_{(\psi )})+\sqrt{-1}\func{Im}(z_{(\psi
)}).
\end{equation}
The new metric associated with such a deformation is provided by 
\begin{equation}
\left| dz_{(\psi )}^{\prime }\right| ^{2}=\frac{(K+1)^{2}}{4}\left|
dz_{(\psi )}+\frac{K-1}{K+1}d\overline{z}_{(\psi )}\right| ^{2}.
\end{equation}
\ Since $dz_{(\psi )}^{2}$ is the given quadratic differential $\psi =\mu
(\zeta )d\zeta \otimes d\zeta $, we can equivalently write $\left| dz_{(\psi
)}^{\prime }\right| ^{2}$ as 
\begin{equation}
\left| dz_{(\psi )}^{\prime }\right| ^{2}=\frac{(K+1)^{2}}{4}\left| \mu
\right| \left| d\zeta +\frac{K-1}{K+1}\left( \frac{\overline{\mu }}{\left|
\mu \right| ^{1/2}}\right) d\overline{\zeta }\right| ^{2},
\end{equation}
where $(\overline{\mu }/\left| \mu \right| )$\ $d\overline{\zeta }\otimes
d\zeta ^{-1}$ is the (\ Teichm\"{u}ller-)\ Beltrami form associated with the
quadratic differential $\psi $.\ \ If we consider quadratic differentials $%
\psi =\mu (\zeta )d\zeta \otimes d\zeta $ in the open unit ball $%
Q_{N_{0}}^{(1)}(M)\doteq \{\psi |,\;||\mu (\zeta )||<1\}$ in the
Teichm\"{u}ller norm\ (\ref{teichnorm}), then there is a\ natural choice for
the constant $K$ provided by 
\begin{equation}
K=\frac{1+||\mu (\zeta )||}{1-||\mu (\zeta )||}.
\end{equation}
In this latter case we get 
\begin{equation}
\left| dz_{(\psi )}^{\prime }\right| ^{2}=\frac{||\mu (\zeta )||^{2}}{(||\mu
(\zeta )||-1)^{2}}\left| \mu \right| \left| d\zeta +\frac{1}{||\mu (\zeta )||%
}\left( \frac{\overline{\mu }}{\left| \mu \right| ^{1/2}}\right) d\overline{%
\zeta }\right| ^{2}.  \label{slice}
\end{equation}
According to Teichm\"{u}ller's existence theorem any complex structure on
can be parametrized by the metrics (\ref{slice}) as $\psi =\mu (\zeta
)d\zeta \otimes d\zeta $, varies in $Q_{N_{0}}^{(1)}(M)$. This is equivalent
to saying that for any given $(M,g)$, (with $(M,N_{0};[g])$ defining a
reference complex structure $\mathcal{C}_{0}$ on $(M,N_{0})$), and any
diffeomorphism $f\in Diff_{0}(M;N_{0})$ mapping $(M,g)$ into $(M,g_{1})$,
with $(M,N_{0};[g_{1}])$ a complex structure distinct from $(M,N_{0};[g])$,
there is a quadratic differential $\psi \in Q_{N_{0}}^{(1)}(M)$ and a
biholomorphic map $F\in Diff_{0}(M;N_{0})$, homotopic to $f$ such that $%
[F^{\ast }g_{1}]$ is given by the conformal class associated with (\ref
{slice}). This is the familiar point of view which allows to identify
Teichm\"{u}ller space with the open unit ball $Q_{N_{0}}^{(1)}(M)$ in the
space of quadratic differentials $Q_{N_{0}}(M)$. It is also worthwhile
noticing that (\ref{slice}) allows us to consider the open unit ball $%
Q_{N_{0}}^{(1)}(M)$ in the space of quadratic differential as providing a
slice for the combined action of $Diff_{0}(M;N_{0})$ and of the conformal
group $W^{s}(M;N_{0})$ on the space of Riemannian metrics $Riem(M;N_{0})$, 
\emph{i.e.} 
\begin{gather}
Q_{N_{0}}^{(1)}(M)\hookrightarrow \frac{Riem(M;N_{0})}{Diff_{0}(M;N_{0})}%
\simeq W^{s}(M;N_{0})\times \mathfrak{T}_{g,N_{0}}(M)  \label{confslice} \\
\left[ \left| dz_{\psi }^{\prime }\right| ^{2}\right] \longmapsto \frac{%
||\mu ||^{2}}{(||\mu ||-1)^{2}}\left| \mu \right| \cdot \left| d\zeta +\frac{%
1}{||\mu ||}\left( \frac{\overline{\mu }}{\left| \mu \right| ^{1/2}}\right) d%
\overline{\zeta }\right| ^{2},  \notag
\end{gather}
where 
\begin{equation}
W^{s}(M;N_{0})\doteq \left\{ f:M\rightarrow \mathbb{R}^{+}|\;f\in H^{s}(M,%
\mathbb{R})\right\}
\end{equation}
denotes the (Weyl) space of conformal factors defined by of all positive
(real valued) functions on $M$ whose derivatives\ up to the order $s$ exist
in the sense of distributions and are represented by square integrable
functions.\ 

\noindent Together with $Q_{N_{0}}(M)$ one introduces also the space $B_{N_{0}}(M)$\
of \ ($L^{\infty }$ measurable) Beltrami differentials, \emph{i.e.} of
tensor fields $\omega =\nu (\zeta )d\overline{\zeta }\otimes d\zeta ^{-1}$,
sections of $k^{-1}\otimes \overline{k}$, ($k$ being the holomorphic
cotangent bundle to $M$), with $\sup_{M}|\nu (\zeta )|<\infty $. \ \ The
space of Beltrami differentials is naturally identified with the tangent
space to $\mathfrak{T}_{g,N_{0}}(M)$, \emph{i.e.}, $\omega =\nu (\zeta )d%
\overline{\zeta }\otimes d\zeta ^{-1}\in T_{C}\mathfrak{T}_{g,N_{0}}(M)$, (with $%
\mathcal{C}$ a complex structure in $\mathfrak{T}_{g,N_{0}}(M)$).

The two spaces $Q_{N_{0}}(M)$ and $B_{N_{0}}(M)$ can be naturally paired
according to 
\begin{equation}
\left\langle \psi |\omega \right\rangle =\int_{M}\mu (\zeta )\nu (\zeta
)d\zeta d\overline{\zeta }.  \label{pair}
\end{equation}
In such a sense $Q_{N_{0}}(M)$ \ is $\mathbb{C}$-anti-linear isomorphic to$\
T_{C}\mathfrak{T}_{g,N_{0}}(M)$, and \ can be canonically identified with the
cotangent space $T_{C}^{\ast }\mathfrak{T}_{g,N_{0}}(M)$ to $\mathfrak{T}%
_{g,N_{0}}(M)$. On the cotangent bundle $T_{C}^{\ast }\mathfrak{T}_{g,N_{0}}(M)$
we can define the Weil-Petersson metric as the inner product between
quadratic differentials corresponding to the $L^{2}$- norm provided by 
\begin{equation}
\left\| \psi \right\| _{WP}^{2}\doteq \int_{M}h^{-2}(\zeta )\left| \psi
(\zeta )\right| ^{2}\left| d\zeta \right| ^{2},
\end{equation}
where $\psi \in Q_{N_{0}}(M)$ and $h(\zeta )\left| d\zeta \right| ^{2}$ is
the hyperbolic metric on $M$ disk-wise described by (\ref{hyper}). Note that 
$\frac{\psi }{h}$ is a Beltrami differential on $M$, thus if we introduce a
basis $\{\mu _{\alpha }\}_{\alpha =1}^{3g-3+N_{0}}$ of the vector space of
harmonic Beltrami differentials on $(M,N_{0})$, we can write 
\begin{equation}
G_{\alpha \overline{\beta }}=\int_{M}\mu _{\alpha }\overline{\mu _{\beta }}%
h(\zeta )\left| d\zeta \right| ^{2}  \label{WPmet}
\end{equation}
for the components of the Weil-Petersson metric on the tangent space to $%
\mathfrak{T}_{g,N_{0}}(M)$. We can introduce the corresponding Weil-Petersson
K\"{a}hler form according to 
\begin{equation}
\omega _{WP}=\sqrt{-1}G_{\alpha \overline{\beta }}dZ^{\alpha }\wedge d%
\overline{Z^{\beta }}.  \label{W-P}
\end{equation}
Such K\"{a}hler potential can be made invariant under the mapping class
group $\mathfrak{Map}(M;N_{0})$ to the effect that the Weil-Petersson volume $2$%
-form $\omega _{WP}$ on $\mathfrak{T}_{g,N_{0}}(M)$ descends on $\mathfrak{M}%
_{g},_{N_{0}}$, and it has a (differentiable) extension, in the sense of
orbifold, to $\overline{\mathfrak{M}}_{g},_{N_{0}}$.

\section{\protect\bigskip Appendix II : relations with curvature assignments
enumeration techniques}

\bigskip 

To the attentive reader it will not have escaped the fact that the statement
of  proposition 12 seems partly in contrast with some of the results
discussed in our monograph [34], where $\gamma _{g}=\frac{5g-1}{2}$ is to
some extent related to a quantity resembling $Card\left[ \{q(i)%
\}_{i=1}^{N_{0}}\right] $. The origin of such a contrast is formula (\ref
{factor}) which is similar to formula 5.33 (p. 103) of [34]. There, one
introduces a quantity $p_{\lambda }^{curv}$, (where $\lambda =N_{0}$ in the
present notation), which plays the role of $Card\left[ \{q(i)\}_{i=1}^{N_{0}}%
\right] $ and which is related to the number of distinct (unlabelled)
curvature assignments over the $N_{0}$ vertices. $p_{\lambda }^{curv}$ is
actually combinatorially estimated by a rather delicate procedure which
calls into play a rooting (at a curvature blob). Note in particular that
strictly speaking while $p_{\lambda }^{curv}$ is obtained by an unlabelled
rooted counting (using partition of integers techniques), \ $Card\left[
\{q(i)\}_{i=1}^{N_{0}}\right] $ is actually obtained by a labelled
enumeration with the labellings formally factorized out by dividing by $%
N_{0}!$, (already in such a sense it is difficult to compare the subleading
asymptotics of the two quantities, since, even if it is true that by dividing
by $N_{0}!$ one removes\ the labels, there are necessary conditions on the
large $N_{0}$ limit of $Card\left[ \{q(i)\}_{i=1}^{N_{0}}\right] $ in order
to carry a subdominant asymptotic comparison with $p_{\lambda }^{curv}$.
Typical example in such a direction are provided by labelled and unlabelled
counting of graphs, see \emph{e.g.} [35] section 9.3). At any rate, this is
not the real source of ambiguity in comparing our results  with those
described in [34].  In 5.33 it further appears the factor $%
<Card\{T_{a}^{(i)}\}_{curv}>$ which in its role it is analogous to $%
VOL\left( \overline{\mathfrak{M}}_{g},_{N_{0}}\right) $, and which must be
normalized by a modular part (see chapter 4 of [34]), in order to provide
the final triangulation counting formula 6.22 of [34]. As a matter of fact,
the estimation of $<Card\{T_{a}^{(i)}\}_{curv}>$ in [34] requires that one
should specify also a rooting fixing the modular degrees of freedom of the
possible deformations of the dynamical triangulations. By specifying the
dimension to $n=2$ and by considering the case of the sphere, where modular
degrees of freedom in the sense of [34] are absent, 6.22 shows that the
corresponding $\gamma _{g=2}=-\frac{1}{2}$ originates from $p_{\lambda
}^{curv}$ which seems at variance with the results of the present analysis.
However, in the present case, also for the sphere, triangulations in $%
\mathcal{DT}[\{q(i)\}_{i=1}^{N_{0}}]$ do possess modular degrees of freedom
(since the counting procedure involving the use of $VOL\left( \overline{%
\mathfrak{M}}_{g},_{N_{0}}\right) $ maps a triangulated sphere into a punctured
sphere). Thus, the correct dictionary between (\ref{factor}) and 5.33 and
6.22 of [34], in the case of the sphere, is that one should compare $%
p_{\lambda }^{curv}$ with $Card\left[ \{q(i)\}_{i=1}^{N_{0}}\right]
VOL\left( \overline{\mathfrak{M}}_{0},_{N_{0}}\right) $, a comparison which
indeed provides the correct interplay between these two rather different
formalisms. Further geometrical aspects clarifying the role of the
Weil-Petersson volume in simplicial quantum gravity are discussed in [36].

\bigskip 

\bigskip 

\noindent\textbf{Acknowledgements}

The authors wish to express their gratitude to C. Dappiaggi for many invaluable
comments. This work has been supported in part by the Ministero dell'Universita' e della
Ricerca Scientifica under the PRIN project \emph{The geometry of integrable
systems.}

\end{document}